\documentclass[a4paper]{article}
\usepackage[affil-it]{authblk}

\usepackage{amsmath}
\usepackage{amssymb}
\usepackage{tikz}
\usepackage{graphicx}
\usepackage[footnotesize,hang]{caption}
\usepackage{pstricks}
\usepackage{subfig}
\usepackage{color}
\usepackage{rotating}
\usepackage{mathtools}
\usepackage{xcolor}
\usepackage{pifont}

\usetikzlibrary{decorations.pathmorphing}

\newcommand{\eps}{\epsilon}

\newcommand{\e}{\mathrm{e}}

\newcommand{\diff}[2]{\frac{\mathrm{d} #1}{\mathrm{d} #2}}
\newcommand{\PL}{Painlev\'e }
\newcommand{\Tq}{tronqu\'ee}
\newcommand{\Ttq}{tritronqu\'ee}

\def\XXint#1#2#3{{\setbox0=\hbox{$#1{#2#3}{\int}$}
\vcenter{\hbox{$#2#3$}}\kern-.5\wd0}}

\usepackage{setspace}
\title{Stokes Phenomena in Discrete Painlev\'e I}
\author{N. Joshi and C. J. Lustri%
  \thanks{Electronic address: \texttt{christopher.lustri@sydney.edu.au}; Corresponding author}}
\affil{School of Mathematics and Statistics, F07, The University of Sydney, New South Wales 2006, Australia}
\date{}                                          

\numberwithin{equation}{section}
\numberwithin{figure}{section}

\begin{document}
\maketitle

\begin{abstract}
In this study, we consider the asymptotic behaviour of the first discrete {\PL}equation in the limit as the independent variable becomes large. Using an asymptotic series expansion, we identify two types of solutions which are pole-free within some sector of the complex plane containing the positive real axis. Using exponential asymptotic techniques, we determine the Stokes Phenomena effects present within these solutions, and hence the regions in which the asymptotic series expression is valid. From a careful analysis of the switching behaviour across Stokes lines, we find that the first type of solution is uniquely defined, while the second type contains two free parameters, and that the region of validity may be extended for appropriate choice of these parameters. 
\end{abstract}

\section{Introduction}
We consider the asymptotic limit  $n \rightarrow \infty$ of the first discrete \PL equation (dP$_\mathrm{I}$)
\begin{equation}\label{0:dPI}
w_{n+1}+w_n + w_{n-1} = \frac{\alpha n + \beta}{w_n} + \gamma,
\end{equation}
and apply methods of exponential asymptotics to describe solutions which demonstrate asymptotically pole-free behaviour within certain sectors. This nonlinear equation also appears in the study of orthogonal polynomials, known there as the Freud equation \cite{Duits1,Magnus2,Magnus1}. Previous  studies \cite{Bernardo1,Joshi2,Vereshchagin1} considered the far-field behaviour of solutions to dP$_\mathrm{I}$, finding the asymptotic solution in terms of elliptic functions. Joshi \cite{Joshi2} and Vereshchagin \cite{Vereshchagin1} also determined discrete analogues of asymptotically pole-free solutions behaviours by Boutroux for the continuous first \PL equation (P$_\mathrm{I}$). In this paper, we find the full formal series solutions including exponentially-small terms, study the influence of Stokes phenomena on these solutions, and deduce their regions of validity. The resultant asymptotic solutions share several features with the asymptotically pole-free solutions to P$_{\mathrm{I}}$ known as tronqu\'ee and tritronqu\'ee solutions.

Equation \eqref{0:dPI} tends to P$_{\mathrm{I}}$ in the scaling limit $w_n = 1-h(2y(x)-c/2)$,  $x=\sqrt{h} n - b/2 - c^2/24$, with $\alpha = h^{3/2}a$, $\beta = -\gamma + 3+ h^2 b$, $\gamma = 6 + h c$, as $h \rightarrow 0$. P$_{\mathrm{I}}$ is one of the six \PL equations which play a key role in a range of physical problems, in particular arising as universal models in reductions of the nonlinear Schr{\"o}dinger equation \cite{Dubrovin1}, the Korteweg-de Vries equation \cite{Hastings1, Rosales1} and the Boussinesq equation \cite{Clarkson1}, and appear as models in such diverse fields such as statistical mechanics \cite{McCoy1}, nonlinear optics \cite{Giannini1}, electrochemistry \cite{Bass1} and plasma physics \cite{Hastings1}. The first discrete \PL equation also arises in physical applications, such as the study of quantum gravity \cite{Fokas1, Fokas2}.

Motivated by these applications, the study of the Painlev\'e equations in asymptotic limits has been widely pursued. However, the corresponding study of discrete Painlev\'e equations remains sparse. In this paper, we apply techniques of exponential asymptotics for differential-difference and difference equations developed by King and Chapman \cite{King4} in their analysis of the Frenkel-Kontorova model for atomic lattices to dP$_{\mathrm{I}}$, in order to obtain asymptotic approximations which are accurate up to exponentially-small correction terms. 

We note that there have been some previous investigations into exponential asymptotics for difference equations, such as that of \cite{Daalhuis2,Olver1}. The study by Olde Daalhuis \cite{Daalhuis2} performed a Borel-plane analysis to consider the behaviour of second-order linear difference equations and to determine exponentially-improved solutions in terms of an inverse factorial series. In fact, these Borel-plane methods could be applied to the current investigation, and controlled error estimates for the asymptotics could be obtained in this fashion. The primary reason for using the Stokes-smoothing technique described in \cite{King4} is that the analysis may be performed directly on the nonlinear difference equation, rather than requiring the solution to be formulated in terms of an integral expression; this feature permits the analysis to be easily extended to a wide range of more complicated problems.  

\subsection{Background}

The asymptotic behaviour of solutions to the \PL equations was studied by Boutroux \cite{Boutroux1}, who considered the behaviour of the first \PL equation (P$_\mathrm{I}$),
\begin{equation*}\label{0:PI}
\diff{^2y}{x^2} = 6y^2 + x,
\end{equation*}
in the limit $|x| \rightarrow \infty$. Boutroux identified special solutions which are asymptotically free of poles in particular sectors of the complex plane of angular width $4\pi/5$, and furthermore discovered five unique solutions that are pole-free in sectors of angular width $8\pi/5$. These solutions are known as \textit{\Tq} and \textit{\Ttq} solutions respectively, and tend to $w\sim \pm\sqrt{-x/6}$ as $|x| \rightarrow \infty$ within the asymptotically pole-free sectors. In \cite{Joshi3}, the real {\Ttq} solution was shown to have no poles on the whole semi-axis $x < 0$. Several asymptotic studies of P$_\mathrm{I}$ have been carried out in \cite{Joshi1, Kapaev1, Kitaev1}. These asymptotic approximations have played essential roles in many of the physical problems described earlier \cite{Dubrovin1, Bass1}.

\subsection{Discrete Painlev\'e equations}

In this study, we are interested in the asymptotic behaviour of solutions of the first discrete Painlev\'e equation. The discrete Painlev\'e equations are a set of integrable nonlinear difference equations, which tend to the ordinary Painlev\'e equations when the continuum limit is taken.

The discrete Painlev\'e equations first appeared in the study of solutions to the continuous Painlev\'e equations. It was discovered that these equations possess B\"acklund transformations, which map one solution to another solution with different parameter values, or even to a solution of a different Painlev\'e equation (see for example \cite{Clarkson1, Gromak1}). These transformations have been used to construct hierarchies of solutions to the Painlev\'e equations, often in terms of rational or special functions.

It was noted by Fokas, Its and Kitaev \cite{Fokas1,Fokas2} that dP$_{\mathrm{I}}$ arises in the study of B\"acklund transformations associated with the fourth Painlev\'e equation (P$_{\mathrm{IV}}$). This equation has two parameters, $a$ and $b$, and is given by
\begin{equation*}
\diff{^2w}{z^2} = \frac{(w')^2}{2w} + \frac{3w^2}{2} + 4zw^2 + 2(z^2-a)w + \frac{b}{w}.
\end{equation*}  
If we associate a choice of parameters $(a_n, b_n)$ of P$_{\mathrm{IV}}$ with a solution denoted by $w_n(z)$, a hierarchy of solutions may be found corresponding to parameter choices
\begin{equation}\label{0:P4par}
a_n =  - \tfrac{3}{2}\mu(-1)^n + \tfrac{1}{2}n,\qquad \sqrt{-2b_n} =  \mu(-1)^n + n,\\
\end{equation}
where $\mu$ is an arbitrary complex-valued constant. The recurrence relation for $w_n$ is subsequently given by
\begin{equation}\label{0:P4dP1}
(w_{n+1} + w_n + w_{n-1})w_n = -2zw_n - n + \mu,
\end{equation}
which is simply a version of dP$_{\mathrm{I}}$ with appropriately chosen parameters. A variant of this relationship was used by \cite{Bassom1} to describe hierarchies of solutions to P$_{\mathrm{IV}}$ in terms of parabolic cylinder functions, as well as in \cite{Clarkson1} to construct solutions in terms of rational functions, as well as Okamoto and Hermite polynomials. By studying the asymptotic behaviour of the discrete Painlev\'e equations, we are able to better understand the behaviour of these solution families.

The first discrete Painlev\'e equation has also featured in the study of problems arising outside of the field of integrable systems, particularly relating to the study of orthogonal and orthonormal polynomials \cite{Magnus2, Magnus1}. A set of orthonormal polynomials $p_n(x)$, where $n \geq 0$, is defined with respect to some weight function $w(x)$ as satisfying
\begin{equation}\label{0:pnw}
\int_{-\infty}^{\infty} p_n(x) p_m(x) w(x) \mathrm{d} x = \delta_{m,n},
\end{equation}
where $\delta_{m,n} = 1$ if $m = n$ and is equal to zero otherwise. In general, these polynomials satisfy a recurrence relation
\begin{equation*}
x p_n(x) = a_{n+1}p_{n+1}(x) + b_n p_n(x) + a_n p_{n-1}(x),
\end{equation*}
where $a_n$ and $b_n$ themselves satisfy particular recurrence relations, described for example in \cite{Magnus2}. In the case of orthonormal polynomials with respect to the Freud weight, $w(x) = \exp(- \kappa x^4 - \mu x^2)$, we find that $b_n = 0$ (this holds for all even weight functions), and that $u_n = a_n^2$ satisfies
\begin{equation}\label{0:pndPI}
4\kappa u_n(u_{n-1} + u_n + u_{n+1}) + 2\mu u_n - n = 0,
\end{equation}
which, again, is dP$_{\mathrm{I}}$, with appropriately chosen parameters. By analyzing the solutions to dP$_{\mathrm{I}}$, we can determine the behaviour of families of orthogonal polynomials, even in the limit that their degree becomes large.  It is this relationship to orthonormal polynomials that underpins the appearance of dP$_{\mathrm{I}}$ in the study of quantum gravity models \cite{Fokas1, Fokas2}. We also note in passing that the variation of $w_n$ with respect to the parameters $\kappa$ and $\mu$ can be computed directly from the fourth continuous Painlev\'e equation with appropriate choice of parameters \cite{Magnus2}, thus confirming the existence of another connection between dP$_{\mathrm{I}}$ and P$_{\mathrm{IV}}$. 

In recent decades, however, discrete integrable systems such as dP$_{\mathrm{I}}$ have become an object of study in themselves. In part, this is due to these equations being convenient examples of nonlinear difference equations which, due to their integrability, are more tractable than many equations within this class. Hence, they provide a useful testing ground for mathematical techniques intended for wider applicability. 

The primary motivation for studying these equations, however, is to better understand their solutions. These equations are important discretizations of the continuous Painlev\'e equations; their particular significance is due to the fact that they retain the integrability property upon discretization.  They arise as independent entities in applications and possess interesting solutions, whose properties permit us to understand the behaviour of classes of functions, such as the orthonormal polynomials described above. While addressing such questions about all solutions lies beyond the scope of the present study, it motivates our interest in understanding the asymptotic behaviour of the discrete Painlev\'e equations.

\subsection{Exponential asymptotics and Stokes lines}\label{Method}

The formal series we study in this paper are shown to be divergent (see Section \ref{dPI_geq0}) and contain exponentially-small terms that are invisible to conventional asymptotic approaches. The central idea of exponential asymptotic methods is that a divergent asymptotic series, when truncated optimally, produces an error term that is exponentially small in the asymptotic limit \cite{Boyd1}. The problem may then be rescaled to directly determine the behaviour of this remainder term, allowing the exponentially-small component of the solution to be determined separately to the asymptotic series itself. This idea was introduced by Berry \cite{Berry3,Berry1,Berry4}, and Berry and Howls \cite{Berry5}, who used these methods to determine the behaviour of special functions such as the Airy function. 

As we uncover this exponentially-small behaviour, we will discover two varieties of important curves, identified originally by Stokes \cite{Stokes1} and known as \textit{Stokes lines} and \textit{anti-Stokes lines}. These curves relate to the behaviour of exponentially-small components of the solutions. As Stokes lines are crossed, the exponentially-small behaviour experiences a smooth, rapid change in value in the neighbourhood of the curve itself; in many cases, the behaviour may switch off entirely on one side of the Stokes line. Hence, Stokes lines allow us to determine which exponentially-small contributions may be present in certain regions of the plane. Anti-Stokes lines determine directions where exponentially-small behaviour changes to become exponentially large, and therefore mark the boundaries of the region in which the asymptotic series expression for the solution is valid. 

In this investigation, we will apply an exponential asymptotic method developed by Olde Daalhuis \textit{et al.} \cite{Daalhuis1} for linear differential equations, and extended by Chapman \textit{et al.} \cite{Chapman1} for application to nonlinear differential equations. We provide a brief outline of the process here, however more detailed explanation of the methodology may be found in these studies.

To optimally truncate an asymptotic series, we require a general form for the asymptotic series coefficients, which is often algebraically intractable. In singular perturbation problems, Dingle \cite{Dingle1} noted that successive terms in the asymptotic series expansion are obtained by repeated differentiation of an earlier term in the series. Hence, the late-order terms $a_m$ of the resultant asymptotic series typically diverge as the ratio between a factorial and the increasing power of a function $\chi$, such as 
\begin{equation}\label{ch1:ansatz}
a_m \sim \frac{A \Gamma(m+\gamma)}{\chi^{m+\gamma}} \qquad \mathrm{as} \quad m \rightarrow \infty,
\end{equation}
where $\Gamma$ is the gamma function defined in \cite{Abramowitz1}, and  $A$, $\gamma$ and $\chi$ are functions that do not depend on $m$. In particular, the \textit{singulant} $\chi$ is equal to zero when the leading-order behaviour is singular, ensuring that this singularity is present in all higher-order terms. Chapman \textit{et al.} \cite{Chapman1} therefore propose a factorial-over-power ansatz (\ref{ch1:ansatz}) for the late-order terms, which may then be used to optimally truncate the asymptotic expansion.

The next step, described by Olde Daalhuis \textit{et al.} \cite{Daalhuis1}, is to substitute the truncated series expression back into the original problem, obtaining an equation for the exponentially-small remainder term after truncation. The positions of Stokes lines are found by determining the curves along which the remainder term varies rapidly, and the behaviour of this remainder is found by solving the remainder equation in the neighbourhood of these Stokes lines.

Away from the Stokes line, the governing behavior is obtained by solving the homogenous version of the remainder equations, generally giving an expression of the form $A\e^{-\chi/\eps}$ (this is a consequence of the analysis in Chapman \textit{et al.} \cite{Chapman1}, although it is also shown as a part of the Stokes smoothing analysis contained herein). Stokes switching will occur when the switching exponential is maximally subdominant; this occurs when the singulant is purely real and positive.  Anti-Stokes lines may be found when the exponential term switches from being small to large; this occurs when the singulant is purely imaginary. These are therefore convenient conditions to determine the possible location of Stokes and anti-Stokes lines, as observed by Dingle \cite{Dingle1}, and we apply them to the series solutions of dP$_{\mathrm{I}}$.

\subsection{Paper outline}
In Section 2, we find the full formal series expansions of the (pole-free) solutions of dP$_\mathrm{I}$ and provide recurrence relations for the coefficients. It is important to note that we find two types of series solutions, called Type A and B solutions. We also describe the late-order terms in these series. We determine the Stokes structures of the Type A solutions in Section 3, which involves descriptions of the exponentially small terms in the solutions. These are used to find the regions in which the solutions are pole-free. We carry out the analysis of Type B solutions in Section 4. Finally, we discuss the results and conclusions of the paper in Section 5.

\section{Asymptotic series expansions}\label{dPI_geq0}
In this section, we expand the solution as a power series in the limit $n \rightarrow \infty$, and determine a recurrence relation for the coefficients of the series. Finally, we determine a general expression for the late-order terms of the series.  

Consider the first discrete \PL equation:
\begin{equation*}
w_{n+1}+w_n+w_{n-1} = \frac{\alpha n + \beta}{w_n} + \gamma.
\end{equation*}
Now, simply scaling into the far field requires the behaviour to be smooth in this limit, however we also wish to consider asymptotically pole-free solutions that allow for the odd and even terms to vary rapidly. Hence, we consider separately the odd and even terms in the series, such that
\begin{equation}\label{2:yz}
w_n =
\Bigg\{
	\begin{array}{ll}
		 y_n & \mbox{if } n \textrm{ even} \\
		 z_n & \mbox{if } n \textrm{ odd}
	\end{array}
,
\end{equation}
where $y_n$ and $z_n$ do not necessarily have the same asymptotic behaviour as $n \rightarrow \infty$. Equation (\ref{1:PL1a}) may now be represented as a system, giving
\begin{subequations}
\begin{align}\label{1:PL1a}
(z_{n+1}+y_n+z_{n-1})y_n &= {\alpha n + \beta + \gamma y_n}, \qquad n \textrm{ even}\\
(y_{n+1}+z_n+y_{n-1})z_n &= {\alpha n + \beta + \gamma z_n}, \qquad n \textrm{ odd}.
\label{1:PL1b}
\end{align}
\end{subequations}

In subsequent analysis, we assume $n\in{\mathbb C}$ with iteration $n\mapsto n+1$ acting on domains, and takes an initial value as given by a function $f(n)$ in a domain that includes the initial point. This is a standard interpretation \cite{Spiegel1,Watson1}, and underlies the theory of difference calculus. This allows us to analytically continue the solution in the complex $n$-plane.

We now apply the method of multiple scales in terms of the fast variable $n$, and a slow variable $s = \eps n$. We scale $y_n$ and $z_n$ and express them such that
\begin{equation}\label{1:scaling}
y_n = \frac{u(s,\eps)}{\eps^{1/2}}, \qquad z_n = \frac{v(s,\eps)}{\eps^{1/2}},
\end{equation}
which gives the rescaled equations
\begin{subequations}
\begin{align}\label{1:PL1scaleda}
(v(s+\eps,\eps) + u(s,\eps) + v(s-\eps,\eps))u(s,\eps) &= \alpha s + \eps \beta + \eps^{1/2}\gamma u(s,\eps),\\
(u(s+\eps,\eps) + v(s,\eps) + u(s-\eps,\eps))v(s,\eps) &= \alpha s + \eps \beta + \eps^{1/2}\gamma v(s,\eps). \label{1:PL1scaledb}
\end{align}
\end{subequations}

We assume that $u(s,\eps)$ and $v(s,\eps)$ are analytic in the variable $s$, and expand the solutions to (\ref{1:PL1scaleda})-(\ref{1:PL1scaledb}) as a Taylor series in $s$ to give the system
\begin{subequations}
\begin{align}\label{1:PL1expandeda}
\left[u(s,\eps) + \sum_{j=0}^{\infty} \frac{2\eps^{2j} v^{(2j)}(s,\eps)}{(2j)!} \right]u(s,\eps) &= \alpha s + \eps \beta + \eps^{1/2} \gamma u(s,\eps),\\
\left[v(s,\eps) + \sum_{j=0}^{\infty} \frac{2\eps^{2j} u^{(2j)}(s,\eps)}{(2j)!} \right]v(s,\eps) &= \alpha s + \eps \beta + \eps^{1/2}\gamma v(s,\eps). \label{1:PL1expandedb}
\end{align}
\end{subequations}

We now expand the solutions $u(s,\eps)$ and $v(s,\eps)$ as a formal power series in $\eps^{1/2}$, such that
\begin{equation}\label{1:seriesinf}
u(s,\eps) \sim \sum_{m=0}^{\infty} \eps^{m/2} u_m(s),\qquad v(s,\eps) \sim \sum_{m=0}^{\infty} \eps^{m/2} v_m(s).
\end{equation}

Equations (\ref{1:PL1expandeda})-(\ref{1:PL1expandedb}) therefore become
\begin{subequations}
\begin{align*}
\sum_{m=0}^{\infty}\sum_{r=0}^{\infty} \eps^{(m+r)/2}u_m u_r + \sum_{j=0}^{\infty} \frac{2\eps^{2j}}{(2j)!} \sum_{m=0}^{\infty}\sum_{r=0}^{\infty} \eps^{(m+r)/2}u_m v_r^{(2j)} &= \alpha s + \eps \beta + \gamma \sum_{m=0}^{\infty} \eps^{(m+1)/2} u_m(s),\\
\sum_{m=0}^{\infty}\sum_{r=0}^{\infty} \eps^{(m+r)/2}v_m v_r + \sum_{j=0}^{\infty} \frac{2\eps^{2j}}{(2j)!} \sum_{m=0}^{\infty}\sum_{r=0}^{\infty} \eps^{(m+r)/2}v_m u_r^{(2j)} &= \alpha s + \eps \beta  + \gamma \sum_{m=0}^{\infty} \eps^{(m+1)/2} v_m(s).
\end{align*}
\end{subequations}

Matching orders of $\eps$ as $\eps \rightarrow 0$ gives
\begin{subequations}
\begin{alignat*}{2}
\mathcal{O}(\eps^0):& \qquad & &u_0(u_0 + 2 v_0) = \alpha s,\\ & & &v_0(v_0 + 2u_0) = \alpha s,\\
\mathcal{O}(\eps^{1/2}):& \qquad & &2(u_0u_1 + u_0v_1 + u_1v_0) = \gamma u_0,\\ & & &2(v_0v_1 + v_0u_1 + v_1u_0) = \gamma v_0,\\
\mathcal{O}(\eps^1):& \qquad & &u_1^2 +2 u_0 u_2 +2 u_0 v_2 + 2 v_0 u_2 + 2u_1 v_1 = \beta + \gamma u_1,\\ & & &
v_1^2 +2 v_0 v_2 +2 v_0 u_2 + 2 u_0 v_2 + 2v_1 u_1 = \beta + \gamma v_1.
\end{alignat*}
\end{subequations}
This gives two sets of solutions. The first of these is given by
\begin{equation}\label{1:earlyordersa}
u_0 = -v_0 = \pm\sqrt{-\alpha s},\qquad u_1 = v_1 = \frac{\gamma}{2},\qquad  u_2  = -v_2 = \mp \frac{4\beta-\gamma^2}{8\sqrt{-\alpha s}},
\end{equation} 
while the second is
\begin{equation}\label{1:earlyordersb}
u_0 = v_0 = \pm\sqrt{\frac{\alpha s}{3}},\qquad u_1 = v_1 = \frac{\gamma}{6},\qquad u_2 = v_2 = \pm\frac{(12\beta+\gamma^2)\sqrt{3}}{72\sqrt{\alpha s}},
\end{equation} 
where the upper and lower choice of signs correspond. In subsequent analysis, we will refer to solutions with the leading-order behaviour given by (\ref{1:earlyordersa}) as Type A, while those with leading-order behaviour given by (\ref{1:earlyordersb}) as Type B. 

We see that in both solution types, $u_0$ and $v_0$ contain a singularity at $s=0$, and that this singularity increases in strength by one half in each subsequent term of the asymptotic series. Consequently, we expect that the series will diverge, and exponentially-small Stokes switching behaviour will be present in the solution.

In general, we find
\begin{subequations}
\begin{alignat}{2}\label{1:orderna}
\mathcal{O}(\eps^{m}):& \qquad &\sum_{j=0}^{2m} u_j u_{2m-j} 
                       + \sum_{j=0}^{{\lfloor}m/2{\rfloor}} \frac{2}{(2j)!}\sum_{r=0}^{2m-4j} u_r v^{(2j)}_{2m-r-4j}& = \gamma u_{2m-1},\\
											& & \sum_{j=0}^{2m} v_j v_{2m-j} 
                       + \sum_{j=0}^{{\lfloor}m/2{\rfloor}} \frac{2}{(2j)!}\sum_{r=0}^{2m-4j} v_r u^{(2j)}_{2m-r-4j}& = \gamma v_{2m-1}. 
											\label{1:ordernb}
\end{alignat}
\end{subequations}

In order to determine the optimal truncation point, we must determine the form of the terms in the asymptotic series as $n \rightarrow \infty$. We therefore follow \cite{Chapman1}, and define a late-order ansatz in the limit $n \rightarrow \infty$:
\begin{equation}\label{1:ansatz}
u_{m} \sim \frac{U(s)\Gamma(m/2+k)}{\chi(s)^{m/2+k}},\qquad v_{m} \sim \frac{V(s)\Gamma(m/2+k)}{\chi(s)^{m/2+k}},
\end{equation}
where $\chi$ is the singulant, satisfying $\chi(0) = 0$, $U$ and $V$ are the prefactors, and $k$ is a constant. 

Applying this ansatz to (\ref{1:orderna})-(\ref{1:ordernb}) and matching leading orders of $m$ as $m \rightarrow \infty$ gives
\begin{subequations}
\begin{align}\label{1:lot1a}
2(u_0U + u_0 V + v_0 U) +  2 u_0 V \sum_{j=1}^{\lfloor m/2 \rfloor}\frac{(-\chi')^{2j}}{(2j)!} &= 0,\\
2(v_0V + v_0 U + u_0 V) +  2 v_0 U \sum_{j=1}^{\lfloor m/2 \rfloor}\frac{(-\chi')^{2j}}{(2j)!} &= 0.
\label{1:lot1b}
\end{align}
\end{subequations}
Continuing to the subsequent order as $m \rightarrow \infty$, we find $u_1 = v_1 = \gamma/2$ and $u_1 = v_1 = \gamma/6$ for Type A and B solutions respectively, which are always satisfied. At the next order, we find 
\begin{subequations}
\begin{align}\nonumber
2(u_2U + &u_2 V + v_2 U) + 2u_0 V'\sum_{j=1}^{\lfloor m/2 \rfloor}\frac{(-\chi')^{2j-1}}{(2j-1)!}\\&- 2u_0 V \chi''\sum_{j=1}^{\lfloor m/2 \rfloor}\frac{(-\chi')^{2j-2}}{(2j-1)!}\
\label{1:lota2} + 2u_2 V \sum_{j=1}^{\lfloor m/2 \rfloor-1} \frac{(-\chi')^{2j} }{(2j)!} = 0,
\\
\nonumber 2(v_2V + &v_2 U + u_2 V) + 2v_0 U'\sum_{j=1}^{\lfloor m/2 \rfloor}\frac{(-\chi')^{2j-1}}{(2j-1)!}\\&- 2v_0 U \chi''\sum_{j=1}^{\lfloor m/2 \rfloor}\frac{(-\chi')^{2j-2}}{(2j-1)!}
\label{1:lotb2} + 2v_2 U \sum_{j=1}^{\lfloor m/2 \rfloor-1} \frac{(-\chi')^{2j} }{(2j)!} = 0.
\end{align}
\end{subequations}

To determine the behaviour of the prefactor and singulant terms, we must consider Type A and B solutions separately, and solve (\ref{1:lot1a})-(\ref{1:lot1b}) to find the singulant $\chi(s)$, and then (\ref{1:lota2})-(\ref{1:lotb2}) to find the prefactors $U(s)$ and $V(s)$.

\section{Stokes structure of Type A solutions}\label{CASEA}

In this section, we solve the late-order term equations for Type A solutions. We use the resultant expressions to determine the Stokes structure, as well as the form of the exponentially-small components of the solution. From these results, we determine the region of the complex plane which is asymptotically pole-free. 

\subsection{Late-order terms}\label{Section:LOT}

The singulant equations given in (\ref{1:lot1a})-(\ref{1:lot1b}) reduce to
\begin{equation}\label{LOT:sing}
2 +  2 \sum_{j=1}^{\lfloor m/2 \rfloor}\frac{(-\chi')^{2j}}{(2j)!} = 0.
\end{equation} 
Recalling that we are considering the late-order terms in the limit $m \rightarrow \infty$, the leading-order behaviour of $\chi$ may be found by taking the series to be infinite, introducing error into the singulant behaviour that is only exponentially-small in this limit. This gives
\begin{equation}\label{2:eikonal}
\cosh(-\chi') = 0.
\end{equation}
The solutions to this expression take the form
\begin{equation}\label{2:singulant0}
\chi' = \pm\frac{\mathrm{i}\pi}{2} + M\pi\mathrm{i},
\end{equation}
where $M$ is any integer. In general, the late-order behaviour of $u_n$ and $v_n$ will be the sum of terms associated with the singulants given in (\ref{2:singulant0}), with each term associated with a particular choice of sign and $M$. However, in a similar fashion to the analysis carried out in \cite{King4}, we note that this sum will be dominated in the $n \rightarrow \infty$ limit by the two terms associated with $M = 0$. Hence, we may restrict our attention to these components in our analysis. 

Recalling that $u_0$ and $v_0$ are singular at $s=0$, and hence that $\chi(0) = 0$, we find that the relevant singulants are given by
\begin{equation}\label{2:singulant}
\chi_1 = \frac{\mathrm{i}\pi s}{2},\qquad \chi_2 = -\frac{\mathrm{i}\pi s}{2}.
\end{equation}

To find the prefactor associated with these singulants, we solve (\ref{1:lota2})-(\ref{1:lotb2}). We take the limit of the summation terms as $n \rightarrow \infty$. Equation (\ref{1:lotb2}) becomes
\begin{align*}
2 (v_2 V + u_2 V + v_2 U) + 2v_0 U'\sinh(-\chi') - \frac{2v_0 U \chi''\sinh(-\chi')}{\chi'} + 2v_2 U \left(\cosh(-\chi')-1\right) = 0.
\end{align*}
Applying the form of the singulant given in (\ref{2:singulant0}), we find that this reduces to $U' = 0$. Hence we set the prefactor $U(s) = \Lambda$, where $\Lambda$ takes constant value. Recalling that there are two distinct singulant contributions, we conclude that there are two constants associated with $U$, denoted as $\Lambda_1$ and $\Lambda_2$. A similar analysis performed on (\ref{1:lota2}) produces $V' = 0$. Hence, we find that there are also two constants associated with $V$, which we will denote as $\Lambda_3$ and $\Lambda_4$.

Finally, we note that for the ansatz (\ref{1:ansatz}) to be consistent with the leading-order behaviour given in (\ref{1:earlyordersa}), we must have $k = -1/2$. This gives a singularity of the appropriate strength in the limit $n \rightarrow 0$. Hence, the late-order terms are given by

\begin{equation}\label{2:LOT}
u_m \sim \frac{\Lambda_1\Gamma(m/2-1/2)}{(\mathrm{i}\pi s/2)^{m/2-1/2}} + \frac{\Lambda_2\Gamma(m/2-1/2)}{(-\mathrm{i}\pi s/2)^{m/2-1/2}},\qquad
v_m \sim \frac{\Lambda_3\Gamma(m/2-1/2)}{(\mathrm{i}\pi s/2)^{m/2-1/2}} + \frac{\Lambda_4\Gamma(m/2-1/2)}{(-\mathrm{i}\pi s/2)^{m/2-1/2}}.
\end{equation}

The only component of the late-order terms that remains to be found is the value of the constants $\Lambda_{i}$, which may be determined numerically. We illustrate this on an example problem in Appendix \ref{A_typeA}. 

\subsection{Stokes structure}\label{0:StokesStruct}

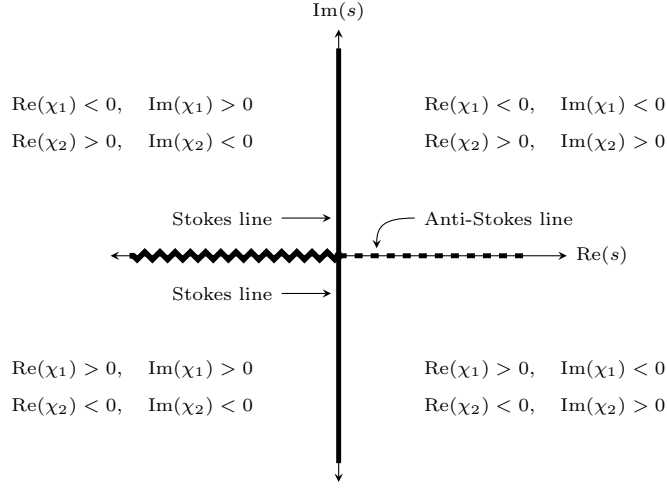
\begin{figure}
\centering
\begin{tikzpicture}
[xscale=0.5,>=stealth,yscale=0.5]

\draw[white] (8,0) -- (-8,0);
\draw[<->] (-6,0) -- (6,0) node[right] {\scriptsize{$\mathrm{Re}(s)$}};
\draw[<->] (0,-6) -- (0,6) node[above] {\scriptsize{$\mathrm{Im}(s)$}};
\draw[line width=0.65mm,dashed] (0,0) -- (5,0);
\draw[line width=0.65mm] (0,0) -- (0,5.5);
\draw[line width=0.65mm] (0,0) -- (0,-5.5);
\draw[decoration = {zigzag,segment length = 2mm, amplitude = 0.5mm},decorate, line width=0.75mm,black] (0,0)--(-5.5,0);
\filldraw (0,0) circle (1mm);

\node at (2,4) [right] {\scriptsize{$\mathrm{Re}(\chi_1) < 0,\quad \mathrm{Im}(\chi_1)<0$}};
\node at (2,3) [right] {\scriptsize{$\mathrm{Re}(\chi_2) > 0,\quad \mathrm{Im}(\chi_2)>0$}};
\node at (2,-3) [right] {\scriptsize{$\mathrm{Re}(\chi_1) > 0,\quad \mathrm{Im}(\chi_1)<0$}};
\node at (2,-4) [right] {\scriptsize{$\mathrm{Re}(\chi_2) < 0,\quad \mathrm{Im}(\chi_2)>0$}};

\node at (-2,4) [left] {\scriptsize{$\mathrm{Re}(\chi_1) < 0,\quad \mathrm{Im}(\chi_1)>0$}};
\node at (-2,3) [left] {\scriptsize{$\mathrm{Re}(\chi_2) > 0,\quad \mathrm{Im}(\chi_2)<0$}};
\node at (-2,-3) [left] {\scriptsize{$\mathrm{Re}(\chi_1) > 0,\quad \mathrm{Im}(\chi_1)>0$}};
\node at (-2,-4) [left] {\scriptsize{$\mathrm{Re}(\chi_2) < 0,\quad \mathrm{Im}(\chi_2)<0$}};

\draw[<-] (-0.2,1) -- (-1.5,1) node[left] {\scriptsize{Stokes line}};
\draw[<-] (-0.2,-1) -- (-1.5,-1) node[left] {\scriptsize{Stokes line}};
\draw[<-] (1,0.2) .. controls (1,0.8) and (1.2,1) .. (2,1) node[right] {\scriptsize{Anti-Stokes line}};

\end{tikzpicture}
\caption{Behaviour of the singulants for Type A solutions in each of the four quadrants. The dashed curve indicates where the real part of both singulants is zero (and hence corresponds to an anti-Stokes line), and the solid curve indicates where the imaginary parts are zero, corresponding to potential Stokes lines. The black jagged line represents a branch cut. Recalling that Stokes switching may only occur when $\mathrm{Im}(\chi) = 0$ and $\mathrm{Re}(\chi) > 0$, we see that $\chi_1$ switches across the curve $\mathrm{arg}(s) = -\pi/2$, while $\chi_2$ switches across the curve $\mathrm{arg}(s) = \pi/2$.}\label{Stokes_dPIa1}
\end{figure}

Having calculated $\chi_1$ and $\chi_2$, we recall from Section \ref{Method} that the behaviour of the exponential term will be proportional to $\exp{(-\chi/\eps)}$; this is also demonstrated in the full Stokes switching analysis contained within Appendix \ref{A2}. This term will be exponentially small when $\mathrm{Re}(\chi) > 0$, and exponentially large when $\mathrm{Re}(\chi) < 0$. We also recall that Stokes lines exist where $\mathrm{Im}(\chi) = 0$ and $\mathrm{Re}(\chi) > 0$, while anti-Stokes lines follow $\mathrm{Re}(\chi) = 0$. Figure \ref{Stokes_dPIa1} illustrates the behaviour of the singulants in each of the four quadrants within the complex $s$-plane.

We see that there is an anti-Stokes line following the positive real axis, while Stokes lines follow the negative and positive imaginary axis for $\chi_1$ and $\chi_2$ respectively. To determine the behaviour that occurs as the Stokes lines are crossed, we note that the remainder term is of algebraic order along anti-Stokes lines. However, this would make it comparable in size to the leading-order behaviour on the positive real axis, which is not possible. Hence, we conclude that the remainder terms associated with $\chi_1$ and $\chi_2$ are not present along the real axis at all, and hence cannot be active in the first or fourth quadrant.

\begin{figure}
\centering
\subfloat[Regions containing exponential behaviour]{
\begin{tikzpicture}
[xscale=0.4,>=stealth,yscale=0.4]

\draw[white] (8,0) -- (-8,0);
\draw[<->] (-6,0) -- (6,0) node[right] {\scriptsize{$\mathrm{Re}(s)$}};
\draw[<->] (0,-6) -- (0,6) node[above] {\scriptsize{$\mathrm{Im}(s)$}};
\draw[line width=0.65mm,dashed] (0,0) -- (5,0);
\draw[line width=0.65mm] (0,0) -- (0,5.5) node[left] {\scriptsize{\ding{193}}};
\draw[line width=0.65mm] (0,0) -- (0,-5.5) node[left] {\scriptsize{\ding{192}}};
\draw[decoration = {zigzag,segment length = 2mm, amplitude = 0.5mm},decorate, line width=0.75mm,black] (0,0)--(-5.5,0);
\filldraw (0,0) circle (1mm);

\draw[<-] (-0.2,1) -- (-1.5,1) node[left] {\scriptsize{Stokes line}};
\draw[<-] (-0.2,-1) -- (-1.5,-1) node[left] {\scriptsize{Stokes line}};
\draw[<-] (1,0.2) .. controls (1,0.8) and (1.2,1) .. (2,1) node[right] {\scriptsize{Anti-Stokes line}};

\node at (2,4) [right] {\scriptsize{$\chi_1$: Inactive}};
\node at (2,3) [right] {\scriptsize{$\chi_2$: Inactive}};
\node at (2,-3) [right] {\scriptsize{$\chi_1$: Inactive}};
\node at (2,-4) [right] {\scriptsize{$\chi_2$: Inactive}};

\node at (-8.5,4) [right] {\scriptsize{$\chi_1$: Inactive}};
\node at (-8.5,3) [right] {\scriptsize{$\chi_2$: Active (exp. small)}};
\node at (-8.5,-3) [right] {\scriptsize{$\chi_1$: Active} (exp. small)};
\node at (-8.5,-4) [right] {\scriptsize{$\chi_2$: Inactive}};

\end{tikzpicture}
}
\subfloat[Exponential contribution]{
\begin{tikzpicture}
[xscale=0.4,>=stealth,yscale=0.4]

\draw[white] (8,0) -- (-8,0);
\draw[<->] (-6,0) -- (6,0) node[right] {\scriptsize{$\mathrm{Re}(s)$}};
\draw[<->] (0,-6) -- (0,6) node[above] {\scriptsize{$\mathrm{Im}(s)$}};
\draw[line width=0.65mm,dashed] (0,0) -- (5,0);
\draw[line width=0.65mm] (0,0) -- (0,5.5) node[left] {\scriptsize{\ding{193}}};
\draw[line width=0.65mm] (0,0) -- (0,-5.5) node[left] {\scriptsize{\ding{192}}};
\draw[decoration = {zigzag,segment length = 2mm, amplitude = 0.5mm},decorate, line width=0.75mm,black] (0,0)--(-5.5,0);
\filldraw (0,0) circle (1mm);

\node at (2,4) [right] {\scriptsize{$R_N = 0$}};
\node at (2,3) [right] {\scriptsize{$S_N = 0$}};
\node at (2,-3) [right] {\scriptsize{$R_N = 0$}};
\node at (2,-4) [right] {\scriptsize{$S_N = 0$}};

\node at (-7.5,4) [right] {\scriptsize{$R_N \sim 4\Lambda_2(\mathrm{i})^{s/\eps}$}};
\node at (-7.5,3) [right] {\scriptsize{$S_N \sim 4\Lambda_4(\mathrm{i})^{s/\eps}$}};
\node at (-7.5,-3) [right] {\scriptsize{$R_N \sim 4\Lambda_1(-\mathrm{i})^{s/\eps}$}};
\node at (-7.5,-4) [right] {\scriptsize{$S_N \sim 4\Lambda_3(-\mathrm{i})^{s/\eps}$}};

\end{tikzpicture}
}
\caption{Figure (a) illustrates the regions of the $s$-plane in which exponential contributions associated with $\chi_1$ and $\chi_2$ are active. Switching behaviour of exponential contributions occurs across the Stokes lines, denoted as solid black lines, where the associated singulant is indicated by the nearby circled numbers. The anti-Stokes line is represented as a dashed curve. The exponential contributions cannot be present along the anti-Stokes line, as they would not be exponentially small here, and hence would be visible to the algebraic power series. Figure (b) shows the exponentially-small remainder terms present in each region.}\label{Stokes_dPIa2}
\end{figure}
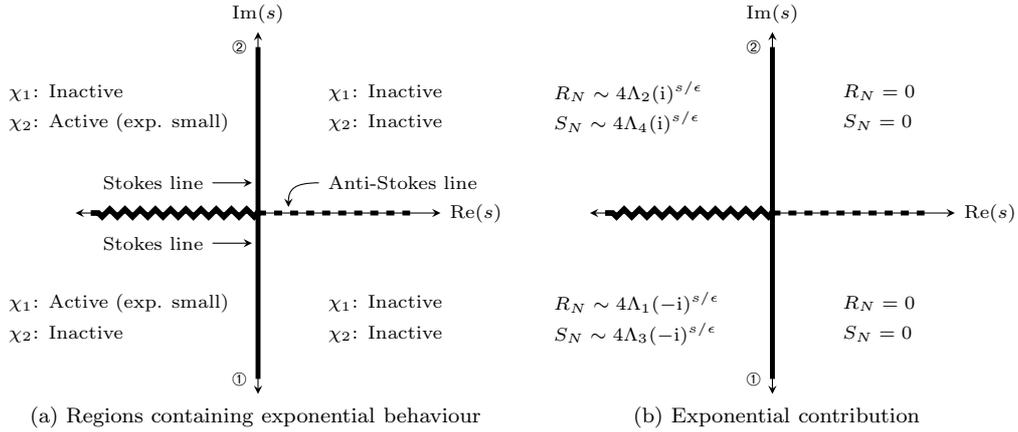

Consequently, we conclude that the exponential contributions associated with $\chi_1$ and $\chi_2$ are present on the left-hand side of their Stokes lines, but are switched off as they cross from the left to the right-hand side (into the fourth and first quadrant respectively). Figure \ref{Stokes_dPIa2} illustrates the regions in which the exponential contributions are active.

Importantly, we see that as the exponential contributions associated with $\chi_1$ and $\chi_2$ are present only in the third and second quadrants respectively, in which they are exponentially small (as $\mathrm{Re}(\chi) > 0$), they do not cause any difficulty for the validity of the asymptotic expansion. Hence, the solutions satisfying (\ref{1:earlyordersa}) are valid on the entire complex plane, with the exception of the branch cut. We therefore conclude that the solutions are asymptotically pole-free in the entire far-field, aside from the negative real axis.

Finally, now that we know the nature of the switching behaviour as the Stokes lines are crossed, we are in a position to calculate the precise form of the remainder term associated with the late-order terms, and hence determine the exponentially-small contribution present in the problem.

\subsection{Stokes smoothing}\label{2.4:Smoothing}

To determine the behaviour in the neighbourhood of the Stokes line, we return to (\ref{1:PL1expandeda})-(\ref{1:PL1expandedb}). Now, we express $u$ and $v$ as asymptotic expansions truncated after $N$ terms, with remainder $R_N$ and $S_N$ respectively, such that
\begin{equation}\label{4:seriesN}
u(s,\eps) = \sum_{m=0}^N \eps^{m/2} u_m(s) + R_N(s),\qquad v(s,\eps) = \sum_{m=0}^N \eps^{m/2} v_m(s) + S_N(s).
\end{equation}
If $N$ is chosen so that the series are truncated after their smallest terms, the remainder terms will be exponentially small. By considering the behaviour of these remainder terms, we can obtain the exponentially-small corrections to the asymptotic series, and use these to determine the region in which the asymptotic series provides a valid approximation of the exact solution.

A full analysis of the remainder terms $R_N$ and $S_N$ is given in Appendix \ref{A2}. This analysis is quite technical; we therefore summarize the results in the remainder of this section. 

From Appendix \ref{A2}, we find that the optimal truncation point is given by $N_{\mathrm{opt}} = 2|\chi|/\eps + \omega$, where $0 \leq \omega < 1$ is chosen so that $N$ is integer-valued. The remainder terms given in (\ref{4:RNout})-(\ref{4:RNout2}) are shown to take the form 
\begin{align}\label{4:remainderR}
R_N    &\sim  \mathcal{S} _1\Lambda_1\,\e^{-\mathrm{i}\pi s/2\eps} + \mathcal{S} _2\Lambda_2\,\e^{\mathrm{i}\pi s/2\eps} = \mathcal{S} _1\Lambda_1\,(-\mathrm{i})^{s/\eps} + \mathcal{S} _2\Lambda_2\,(\mathrm{i})^{s/\eps},\\
S_N    &\sim  \mathcal{S} _1\Lambda_3\,\e^{-\mathrm{i}\pi s/2\eps} + \mathcal{S} _2\Lambda_2\,\e^{\mathrm{i}\pi s/2\eps} = \mathcal{S} _1\Lambda_3\,(-\mathrm{i})^{s/\eps} + \mathcal{S} _2\Lambda_4\,(\mathrm{i})^{s/\eps}.\label{4:remainderS}
\end{align}
where $\mathcal{S} _i$ is the Stokes-switching parameter that varies rapidly in the neighbourhood of Stokes lines. From the analysis in Appendix \ref{A2}, we find that the value of $\mathcal{S}_1$ changes smoothly from $\mathcal{S}_1 = 4$ to $\mathcal{S}_1 = 0$ when crossing from the third to the fourth quadrant, and the value of $\mathcal{S}_2$ changes smoothly from $\mathcal{S}_2 = 0$ to $\mathcal{S}_2 = 4$ when crossing from the second to the first quadrant. Hence, the exponential behaviour of the asymptotic approximation has been completely determined, and is shown in Figure \ref{Stokes_dPIa2}(b).

Consequently, the optimally-truncated asymptotic series for $u(s,\eps)$ and $v(s,\eps)$ takes the form
\begin{align}\label{4:asymp1app}
u(s,\eps) &\sim \sum_{m=0}^{N_{\mathrm{opt}}} \eps^{m/2} u_m(s) + \mathcal{S} _1\Lambda_1\,(-\mathrm{i})^{s/\eps} + \mathcal{S} _2\Lambda_2\,(\mathrm{i})^{s/\eps},\\
\label{4:asymp1app2} v(s,\eps) &\sim \sum_{m=0}^{N_{\mathrm{opt}}}\eps^{m/2} v_m(s) + \mathcal{S} _1\Lambda_3\,(-\mathrm{i})^{s/\eps} + \mathcal{S} _2\Lambda_4\,(\mathrm{i})^{s/\eps},
\end{align}
where the leading orders of $u_n$ and $v_n$ are give in (\ref{1:earlyordersa}), and the late-order behaviour is given in (\ref{2:LOT}). This expression is therefore an accurate asymptotic approximation up to exponentially-small terms, valid in the entire complex $s$-plane.

\section{Stokes structure of Type B solutions}\label{CASEB}

In this section, we determine the late-order asymptotic terms associated with Type B. These are then used to determine the exponentially-small contributions to the asymptotic solution, and subsequently the Stokes structure. From these results, we determine the region of validity for the general solution, and show that for particular choices of the two Stokes-switching parameters, the region of validity may be extended.

\subsection{Late-order terms}
%
%

When considering Type B solutions, we find that the asymptotic series expansion for $u$ and $v$ are identical, and hence that $u_m = v_m$ for all values of $m$. We therefore set $V(s) = U(s)$.

Now, using (\ref{1:earlyordersb}), both singulant equations (\ref{1:lot1a})-(\ref{1:lot1b}) reduce to
\begin{align}\label{1:eikonal}
2 + \cosh(-\chi') = 0,
\end{align}
where we have again taken the upper bound of the summation terms to be infinite. From the leading-order behaviour (\ref{1:earlyordersb}), we know that the solution is singular at $s=0$, and hence that $\chi = 0$ at $s = 0$. Solving (\ref{1:eikonal}) with this boundary condition gives
\begin{equation*}
\chi = \pm\left[\log(-2-\sqrt{3}) + 2 M \pi \mathrm{i}\right]s.
\end{equation*}
where $M$ is some integer. As in Type A solutions, the late-order behaviour for $u_n$ will take the form of an infinite sum of factorial-over-power ansatz terms, each associated with a unique choice of $\chi$. However, at late orders, this sum will be dominated by the four terms associated with $M=0$ and $M=-1$, which have the smallest absolute value. Hence, the four terms that contribute to the exponential behaviour of the solution are
\begin{alignat*}{2}
\chi_1 &= \log(-2-\sqrt{3})s,&\qquad \chi_2 &= -\log(-2-\sqrt{3})s,\\ \chi_3 &= \log(-2+\sqrt{3})s,&\qquad \chi_4 &= -\log(-2+\sqrt{3})s.
\end{alignat*}
We observe that, unlike Type A, there are four sets of late-order behaviour to be considered in order to understand the Stokes structure of the associated solution. By solving (\ref{1:lota2})-(\ref{1:lotb2}), we again find that $U' = 0$. However, in this case, there are four distinct singulant expressions, and hence four constants associated with the corresponding late-order terms (denoted $\Lambda_1$ - $\Lambda_4$). We also find that, as before, $k = -1/2$.

Hence, we find that the late-order terms may be expressed as
\begin{align}
\nonumber u_m = v_m \sim \frac{\Lambda_1\Gamma(m/2-1/2)}{(\log(-2-\sqrt{3})s)^{m/2-1/2}} &+ \frac{\Lambda_2 \Gamma(m/2-1/2)}{(-\log(-2-\sqrt{3})s)^{m/2-1/2}}\\+&\frac{\Lambda_3 \Gamma(m/2-1/2)}{(\log(-2+\sqrt{3})s)^{m/2-1/2}} + \frac{\Lambda_4 \Gamma(m/2-1/2)}{(-\log(-2+\sqrt{3})s)^{m/2-1/2}},\label{4:LOTcalc}
\end{align}
where $\Lambda_i$ is associated with $\chi_i$. 

The numerical process to determine $\Lambda_i$ is more complicated than that demonstrated in Appendix \ref{1:2Inner}, as we must now compute four values of $\Lambda_i$, rather than two. An example numerical calculation is illustrated in Appendix \ref{A_typeB}.

\subsection{Stokes structure}\label{1.3:Structure}

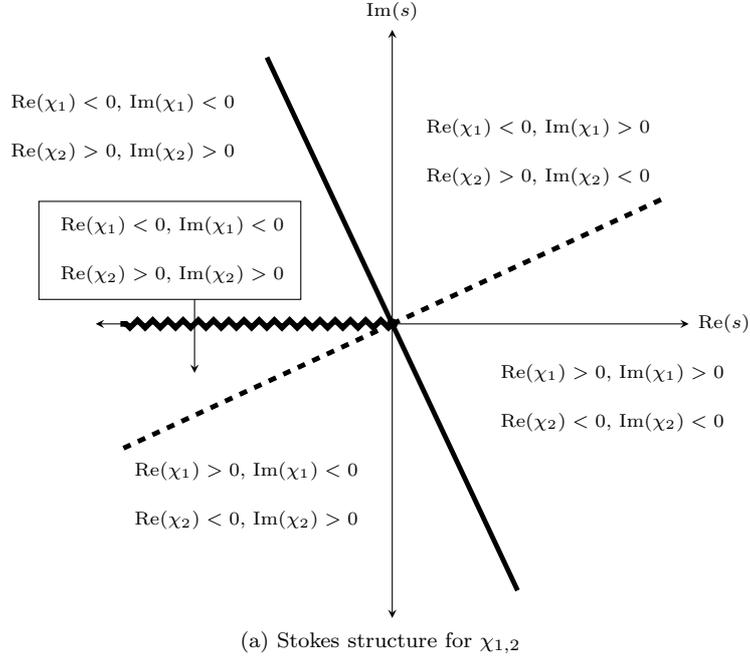
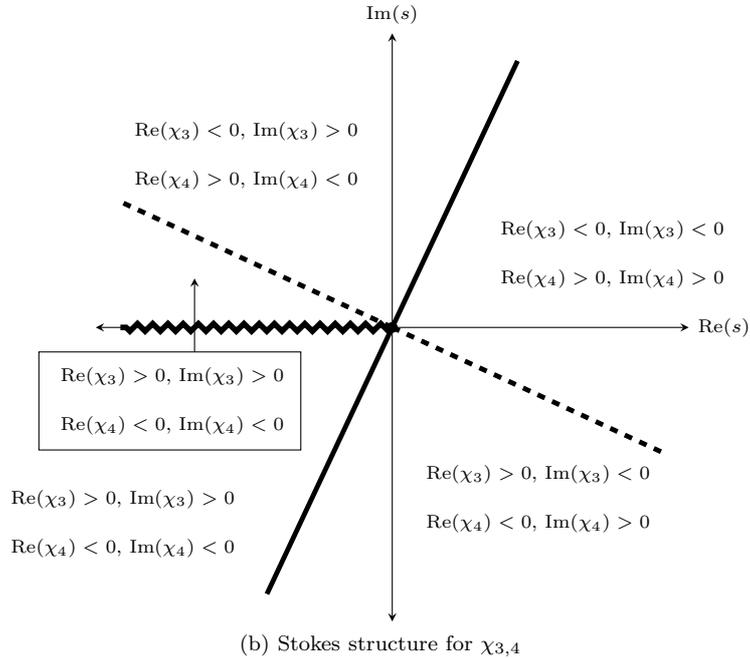
\begin{figure}
\centering

\subfloat[Stokes structure for $\chi_{1,2}$]{
\begin{tikzpicture}
[xscale=0.65,>=stealth,yscale=0.65]

\draw[<->] (-6,0) -- (6,0) node[right] {\scriptsize{$\mathrm{Re}(s)$}};
\draw[<->] (0,-6) -- (0,6) node[above] {\scriptsize{$\mathrm{Im}(s)$}};

\draw[line width=0.65mm,dashed] (0,0) -- (25:6);
\draw[line width=0.65mm] (0,0) -- (-65:6);
\draw[line width=0.65mm] (0,0) -- (25+90:6);
\draw[line width=0.65mm,dashed] (0,0) -- (-65-90:6);

\draw node at (2,-1) [right] {\scriptsize{$\mathrm{Re}(\chi_1) > 0$, $\mathrm{Im}(\chi_1) > 0$}};
\draw node at (2,-2) [right] {\scriptsize{$\mathrm{Re}(\chi_2) < 0$, $\mathrm{Im}(\chi_2) < 0$}};

\draw node at (0.5,4) [right] {\scriptsize{$\mathrm{Re}(\chi_1) < 0$, $\mathrm{Im}(\chi_1) > 0$}};
\draw node at (0.5,3) [right] {\scriptsize{$\mathrm{Re}(\chi_2) > 0$, $\mathrm{Im}(\chi_2) < 0$}};

\draw node at (-2,2) [left] {\scriptsize{$\mathrm{Re}(\chi_1) < 0$, $\mathrm{Im}(\chi_1) < 0$}};
\draw node at (-2,1) [left] {\scriptsize{$\mathrm{Re}(\chi_2) > 0$, $\mathrm{Im}(\chi_2) > 0$}};
\draw [->] (-4,0.5) -- (-4,-1);
\draw (-7.15,2.5) -- (-7.15,0.5) -- (-1.85,0.5) -- (-1.85,2.5) -- cycle;

\draw node at (-0.5,-3) [left] {\scriptsize{$\mathrm{Re}(\chi_1) > 0$, $\mathrm{Im}(\chi_1) < 0$}};
\draw node at (-0.5,-4) [left] {\scriptsize{$\mathrm{Re}(\chi_2) < 0$, $\mathrm{Im}(\chi_2) > 0$}};

\draw node at (-3,4.5) [left] {\scriptsize{$\mathrm{Re}(\chi_1) < 0$, $\mathrm{Im}(\chi_1) < 0$}};
\draw node at (-3,3.5) [left] {\scriptsize{$\mathrm{Re}(\chi_2) > 0$, $\mathrm{Im}(\chi_2) > 0$}};

\draw[decoration = {zigzag,segment length = 2mm, amplitude = 0.5mm},decorate, line width=0.75mm,black] (0,0)--(-5.5,0);

\filldraw (0,0) circle (1mm);
\end{tikzpicture}
}

\subfloat[Stokes structure for $\chi_{3,4}$]{
\begin{tikzpicture}
[xscale=0.65,>=stealth,yscale=0.65]

\draw[<->] (-6,0) -- (6,0) node[right] {\scriptsize{$\mathrm{Re}(s)$}};
\draw[<->] (0,-6) -- (0,6) node[above] {\scriptsize{$\mathrm{Im}(s)$}};

\draw[line width=0.65mm,dashed] (0,0) -- (-25:6);
\draw[line width=0.65mm] (0,0) -- (65:6);
\draw[line width=0.65mm] (0,0) -- (-25-90:6);
\draw[line width=0.65mm,dashed] (0,0) -- (65+90:6);

\draw node at (2,1) [right] {\scriptsize{$\mathrm{Re}(\chi_4) > 0$, $\mathrm{Im}(\chi_4) > 0$}};
\draw node at (2,2) [right] {\scriptsize{$\mathrm{Re}(\chi_3) < 0$, $\mathrm{Im}(\chi_3) < 0$}};

\draw node at (0.5,-4) [right] {\scriptsize{$\mathrm{Re}(\chi_4) < 0$, $\mathrm{Im}(\chi_4) > 0$}};
\draw node at (0.5,-3) [right] {\scriptsize{$\mathrm{Re}(\chi_3) > 0$, $\mathrm{Im}(\chi_3) < 0$}};

\draw node at (-2,-2) [left] {\scriptsize{$\mathrm{Re}(\chi_4) < 0$, $\mathrm{Im}(\chi_4) < 0$}};
\draw node at (-2,-1) [left] {\scriptsize{$\mathrm{Re}(\chi_3) > 0$, $\mathrm{Im}(\chi_3) > 0$}};
\draw [->] (-4,-0.5) -- (-4,1);
\draw (-7.15,-2.5) -- (-7.15,-0.5) -- (-1.85,-0.5) -- (-1.85,-2.5) -- cycle;

\draw node at (-0.5,3) [left] {\scriptsize{$\mathrm{Re}(\chi_4) > 0$, $\mathrm{Im}(\chi_4) < 0$}};
\draw node at (-0.5,4) [left] {\scriptsize{$\mathrm{Re}(\chi_3) < 0$, $\mathrm{Im}(\chi_3) > 0$}};

\draw node at (-3,-4.5) [left] {\scriptsize{$\mathrm{Re}(\chi_4) < 0$, $\mathrm{Im}(\chi_4) < 0$}};
\draw node at (-3,-3.5) [left] {\scriptsize{$\mathrm{Re}(\chi_3) > 0$, $\mathrm{Im}(\chi_3) > 0$}};

\draw[decoration = {zigzag,segment length = 2mm, amplitude = 0.5mm},decorate, line width=0.75mm,black] (0,0)--(-5.5,0);

\filldraw (0,0) circle (1mm);
\end{tikzpicture}
}

\caption{Behaviour of the singulants for Type B solutions in the complex $s$-plane. The dashed lines indicate where the real part of the singulants is zero (and hence corresponds to an anti-Stokes line), and the solid lines indicates where the imaginary parts are zero. The jagged line represents a branch cut. Recalling that Stokes switching may only occur when $\mathrm{Im}(\chi) = 0$ and $\mathrm{Re}(\chi) > 0$, we see that $\chi_2$ and $\chi_4$ switch across the Stokes line in the upper-half plane, while $\chi_1$ and $\chi_3$ switch across the Stokes line in the lower-half plane.}\label{Stokes_dPIab}
\end{figure}

Having calculated the singulants, we may determine the Stokes structure in a similar fashion to  Section \ref{0:StokesStruct}, noting that Stokes lines follow curves on which $\mathrm{Im}(\chi) = 0$ and $\mathrm{Re}(\chi) = 0$, and anti-Stokes lines follow curves satisfying $\mathrm{Re}(\chi) = 0$. The behaviour of the four different singulants is illustrated in Figure \ref{Stokes_dPIab}.

From this figure, we see that the Stokes structure of the solution is substantially more complicated than in Type A solutions. Not only are there four different exponential contributions to consider, but there are two different sets of Stokes lines, each associated with two of the singulants. 

There is, however, also a more interesting difference. From Figure \ref{Stokes_dPIa1}, we were able to conclude that neither exponential contribution could be present on the positive real axis in Type A solutions, as they would not be exponentially small. Hence, as the jump across Stokes lines is prescribed in (\ref{4:remainderR})-(\ref{4:remainderS}), the asymptotic behaviour is specified uniquely, with no free parameters. However, from Figure \ref{Stokes_dPIab}, we see that this is no longer true for Type B solutions. While we can see that $\chi_2$ and $\chi_3$ cannot be present on the positive real axis, as they would be exponentially large in the asymptotic limit, the same cannot be said about $\chi_1$ and $\chi_4$. As $\mathrm{Re}(\chi_{1,4}) > 0$ on the positive real axis, the associated exponential contribution is exponentially small, and the asymptotic series is still valid when such contributions are present. Hence, the value of $\mathcal{S}_{1,4}$ on the real axis may be freely specified, meaning that the solution contains two free parameters.

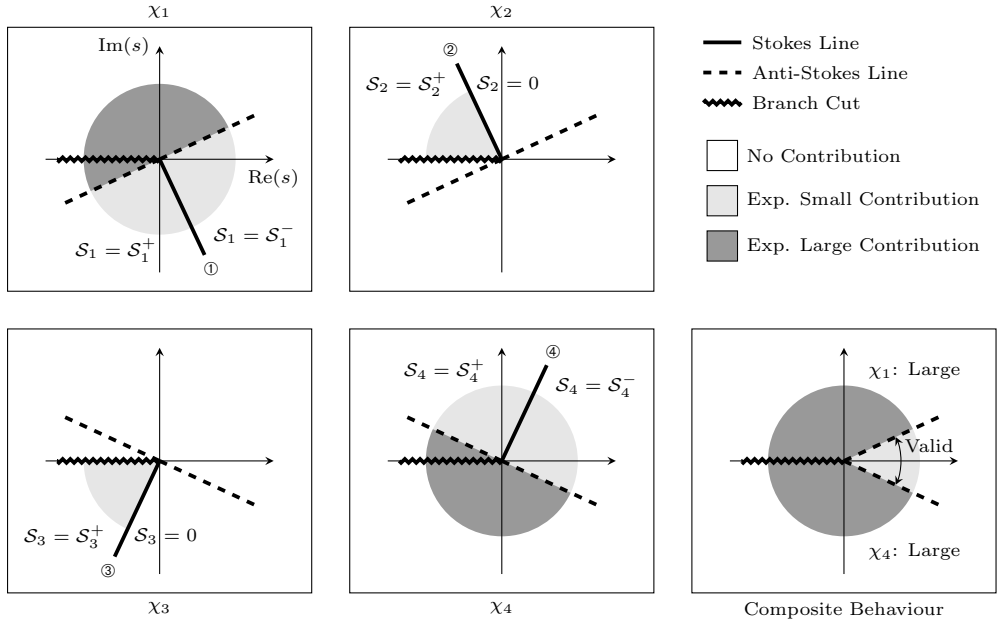
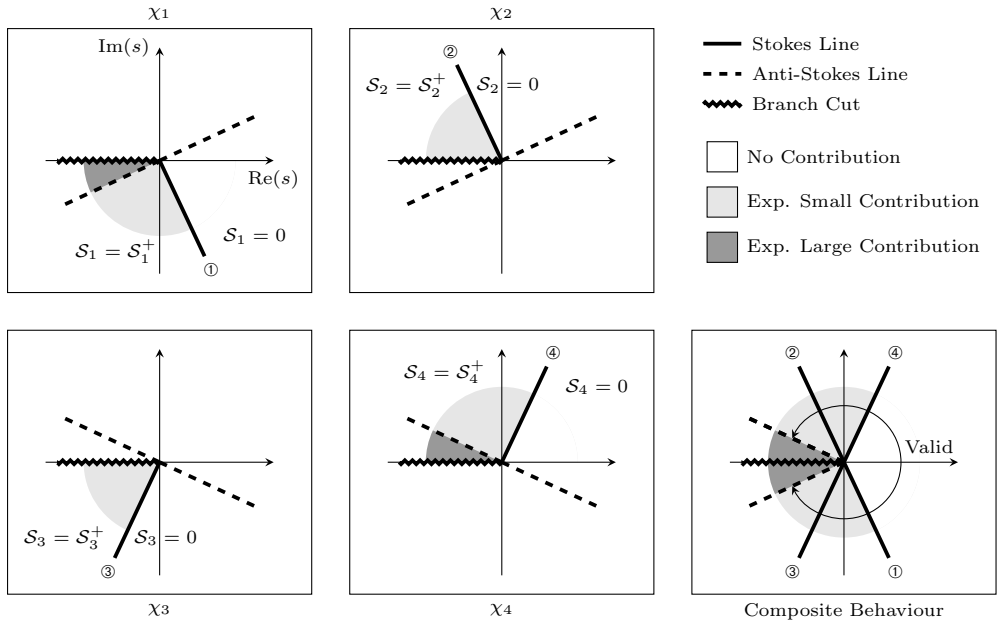
\begin{figure}
\centering

\subfloat[General Stokes Structure]{
\begin{tikzpicture}
[xscale=1,>=stealth,yscale=1]

\draw (0,0) -- (0,3.5) -- (4,3.5) -- (4,0) -- cycle;
\draw (4.5,0) -- (4.5,3.5) -- (8.5,3.5) -- (8.5,0) -- cycle;
\draw (0,4) -- (0,7.5) -- (4,7.5) -- (4,4) -- cycle;
\draw (4.5,4) -- (4.5,7.5) -- (8.5,7.5) -- (8.5,4) -- cycle;
\draw (9,0) -- (9,3.5) -- (13,3.5) -- (13,0) -- cycle;

\draw[line width=0.50mm] (9.15,7.3) -- (9.65,7.3) node[right] {\scriptsize{Stokes Line}};
\draw[line width=0.50mm,dashed] (9.15,6.9) -- (9.65,6.9) node[right] {\scriptsize{Anti-Stokes Line}};
\draw[decoration = {zigzag,segment length = 1mm, amplitude = 0.25mm},decorate, line width=0.5mm,black] (9.15,6.5)--(9.65,6.5) node[right] {\scriptsize{Branch Cut}};
\draw (9.6,5.8) node [right] {\scriptsize{No Contribution}} -- (9.6,6) -- (9.2,6) -- (9.2,5.6) -- (9.6,5.6) -- (9.6,5.8) -- cycle;
\fill[opacity=0.1] (9.2,5.2) -- (9.2,5.4) -- (9.6,5.4) -- (9.6,5) -- (9.2,5) -- (9.2,5.2) -- cycle;
\draw (9.6,5.2) node [right] {\scriptsize{Exp. Small Contribution}}-- (9.6,5.4) -- (9.2,5.4) -- (9.2,5) -- (9.6,5) -- (9.6,5.2) -- cycle;
\fill[opacity=0.4] (9.2,4.6) -- (9.2,4.8) -- (9.6,4.8) -- (9.6,4.4) -- (9.2,4.4) -- (9.2,4.6) -- cycle;
\draw (9.6,4.6) node [right] {\scriptsize{Exp. Large Contribution}}-- (9.6,4.8) -- (9.2,4.8) -- (9.2,4.4) -- (9.6,4.4) -- (9.6,4.6) -- cycle;

\draw[line width=0.50mm,dashed] (2,1.75) -- ++(-25:1.40);
\draw[line width=0.50mm,dashed] (2,1.75) -- ++(-180-25:1.40);
\draw[line width=0.50mm] (2,1.75) -- ++(-25-90:1.40);
\draw[->] (2,0.25) -- (2,3.25);
\draw[->] (0.5,1.75) -- (3.5,1.75);
\draw[decoration = {zigzag,segment length = 1mm, amplitude = 0.25mm},decorate, line width=0.5mm,black] (2,1.75)--(0.65,1.75);
\fill[opacity=0.1] (1,1.75) arc (-180:-90-25:1) -- (2,1.75) -- cycle;
\node at (2,1.75) (a) {};
\path (a) ++(-90-25:1.6) node (c) {\scriptsize{\ding{194}}};
\node at (0.75,0.75)  {\scriptsize{$\mathcal{S}_3 = \mathcal{S}_3^+$}};
\node at (2.08,0.75)  {\scriptsize{$\mathcal{S}_3 = 0$}};

\draw[line width=0.50mm,dashed] (2+4.5,1.75) -- ++(-25:1.40);
\draw[line width=0.50mm,dashed] (2+4.5,1.75) -- ++(-180-25:1.40);
\draw[line width=0.50mm] (2+4.5,1.75) -- ++(-25+90:1.40);
\draw[->] (2+4.5,0.25) -- (2+4.5,3.25);
\draw[->] (0.5+4.5,1.75) -- (3.5+4.5,1.75);
\draw[decoration = {zigzag,segment length = 1mm, amplitude = 0.25mm},decorate, line width=0.5mm,black] (2+4.5,1.75)--(0.65+4.5,1.75);
\fill[opacity=0.4] (1+4.5,1.75) arc (-180:-25:1) -- (2+4.5,1.75) -- cycle;
\fill[opacity=0.4] (1+4.5,1.75) arc (180:180-25:1) -- (2+4.5,1.75) -- cycle;
\fill[opacity=0.1] (3+4.5,1.75) arc (0:-25:1) -- (2+4.5,1.75) -- cycle;
\fill[opacity=0.1] (3+4.5,1.75) arc (0:180-25:1) -- (2+4.5,1.75) -- cycle;
\node at (2+4.5,1.75) (a) {};
\path (a) ++(90-25:1.6) node (c) {\scriptsize{\ding{195}}};
\node at (3.25,0.75+4)  {\scriptsize{$\mathcal{S}_1 = \mathcal{S}_1^-$}};
\node at (1.425,0.55+4)  {\scriptsize{$\mathcal{S}_1 = \mathcal{S}_1^+$}};

\draw[line width=0.50mm,dashed] (2+4.5,1.75+4) -- ++(25:1.40);
\draw[line width=0.50mm,dashed] (2+4.5,1.75+4) -- ++(-180+25:1.40);
\draw[line width=0.50mm] (2+4.5,1.75+4) -- ++(25+90:1.40);
\draw[->] (2+4.5,0.25+4) -- (2+4.5,3.25+4);
\draw[->] (0.5+4.5,1.75+4) -- (3.5+4.5,1.75+4);
\draw[decoration = {zigzag,segment length = 1mm, amplitude = 0.25mm},decorate, line width=0.5mm,black] (2+4.5,1.75+4)--(0.65+4.5,1.75+4);
\fill[opacity=0.1] (1+4.5,1.75+4) arc (180:90+25:1) -- (2+4.5,1.75+4) -- cycle;
\node at (2+4.5,1.75+4) (a) {};
\path (a) ++(90+25:1.6) node (c) {\scriptsize{\ding{193}}};
\node at (0.75+4.5,3.5-0.75+4)  {\scriptsize{$\mathcal{S}_2 = \mathcal{S}_2^+$}};
\node at (2.08+4.5,3.5-0.75+4)  {\scriptsize{$\mathcal{S}_2 = 0$}};

\draw[line width=0.50mm,dashed] (2,1.75+4) -- ++(25:1.40);
\draw[line width=0.50mm,dashed] (2,1.75+4) -- ++(-180+25:1.40);
\draw[line width=0.50mm] (2,1.75+4) -- ++(25-90:1.40);
\draw[->] (2,0.25+4) -- (2,3.25+4) node [left] {\scriptsize{$\mathrm{Im}(s)$}};
\draw[->] (0.5,1.75+4) -- (3.5,1.75+4) node [below] {\scriptsize{$\mathrm{Re}(s)$}};;
\draw[decoration = {zigzag,segment length = 1mm, amplitude = 0.25mm},decorate, line width=0.5mm,black] (2,1.75+4)--(0.65,1.75+4);
\fill[opacity=0.4] (1,1.75+4) arc (180:25:1) -- (2,1.75+4) -- cycle;
\fill[opacity=0.4] (1,1.75+4) arc (180:180+25:1) -- (2,1.75+4) -- cycle;
\fill[opacity=0.1] (3,1.75+4) arc (0:25:1) -- (2,1.75+4) -- cycle;
\fill[opacity=0.1] (3,1.75+4) arc (0:-180+25:1) -- (2,1.75+4) -- cycle;
\node at (2,1.75+4) (a) {};
\path (a) ++(-90+25:1.6) node (c) {\scriptsize{\ding{192}}};
\node at (8.5-0.75,3.5-0.75)  {\scriptsize{$\mathcal{S}_4 = \mathcal{S}_4^-$}};
\node at (8.5-2.75,3.5-0.55)  {\scriptsize{$\mathcal{S}_4 = \mathcal{S}_4^+$}};

\draw[line width=0.50mm,dashed] (2+9,1.75) -- ++(-25:1.40);
\draw[line width=0.50mm,dashed] (2+9,1.75) -- ++(25:1.40);
\draw[->] (2+9,0.25) -- (2+9,3.25);
\draw[->] (0.5+9,1.75) -- (3.5+9,1.75);
\draw[decoration = {zigzag,segment length = 1mm, amplitude = 0.25mm},decorate, line width=0.5mm,black] (2+9,1.75)--(0.65+9,1.75);
\fill[opacity=0.4] (1+9,1.75) arc (-180:-25:1) -- (2+9,1.75) -- cycle;
\fill[opacity=0.4] (1+9,1.75) arc (180:25:1) -- (2+9,1.75) -- cycle;
\fill[opacity=0.1] (3+9,1.75) arc (0:-25:1) -- (2+9,1.75) -- cycle;
\fill[opacity=0.1] (3+9,1.75) arc (0:25:1) -- (2+9,1.75) -- cycle;
\node at (2.2+9,1.2+1.75) [right] {\scriptsize{$\chi_1$: Large}};
\node at (2.2+9,-1.2+1.75) [right] {\scriptsize{$\chi_4$: Large}};
\draw[->] (2+9+0.75,1.75) arc (0:25:0.75); 
\draw[->] (2+9+0.75,1.75) arc (0:-25:0.75);
\node at (2+9+0.675,1.75) [above right] {\scriptsize{Valid}};

\node at (2,0) [below] {\scriptsize{$\chi_3$}};
\node at (6.5,0) [below] {\scriptsize{$\chi_4$}};
\node at (2,7.5) [above] {\scriptsize{$\chi_1$}};
\node at (6.5,7.5) [above] {\scriptsize{$\chi_2$}};
\node at (11,0) [below] {\scriptsize{Composite Behaviour}};


\end{tikzpicture}
}

\subfloat[Special Asymptotic Solution]{
\begin{tikzpicture}
[xscale=1,>=stealth,yscale=1]

\draw (0,0) -- (0,3.5) -- (4,3.5) -- (4,0) -- cycle;
\draw (4.5,0) -- (4.5,3.5) -- (8.5,3.5) -- (8.5,0) -- cycle;
\draw (0,4) -- (0,7.5) -- (4,7.5) -- (4,4) -- cycle;
\draw (4.5,4) -- (4.5,7.5) -- (8.5,7.5) -- (8.5,4) -- cycle;
\draw (9,0) -- (9,3.5) -- (13,3.5) -- (13,0) -- cycle;

\draw[line width=0.50mm] (9.15,7.3) -- (9.65,7.3) node[right] {\scriptsize{Stokes Line}};
\draw[line width=0.50mm,dashed] (9.15,6.9) -- (9.65,6.9) node[right] {\scriptsize{Anti-Stokes Line}};
\draw[decoration = {zigzag,segment length = 1mm, amplitude = 0.25mm},decorate, line width=0.5mm,black] (9.15,6.5)--(9.65,6.5) node[right] {\scriptsize{Branch Cut}};
\draw (9.6,5.8) node [right] {\scriptsize{No Contribution}} -- (9.6,6) -- (9.2,6) -- (9.2,5.6) -- (9.6,5.6) -- (9.6,5.8) -- cycle;
\fill[opacity=0.1] (9.2,5.2) -- (9.2,5.4) -- (9.6,5.4) -- (9.6,5) -- (9.2,5) -- (9.2,5.2) -- cycle;
\draw (9.6,5.2) node [right] {\scriptsize{Exp. Small Contribution}}-- (9.6,5.4) -- (9.2,5.4) -- (9.2,5) -- (9.6,5) -- (9.6,5.2) -- cycle;
\fill[opacity=0.4] (9.2,4.6) -- (9.2,4.8) -- (9.6,4.8) -- (9.6,4.4) -- (9.2,4.4) -- (9.2,4.6) -- cycle;
\draw (9.6,4.6) node [right] {\scriptsize{Exp. Large Contribution}}-- (9.6,4.8) -- (9.2,4.8) -- (9.2,4.4) -- (9.6,4.4) -- (9.6,4.6) -- cycle;

\draw[line width=0.50mm,dashed] (2,1.75) -- ++(-25:1.40);
\draw[line width=0.50mm,dashed] (2,1.75) -- ++(-180-25:1.40);
\draw[line width=0.50mm] (2,1.75) -- ++(-25-90:1.40);
\draw[->] (2,0.25) -- (2,3.25);
\draw[->] (0.5,1.75) -- (3.5,1.75);
\draw[decoration = {zigzag,segment length = 1mm, amplitude = 0.25mm},decorate, line width=0.5mm,black] (2,1.75)--(0.65,1.75);
\fill[opacity=0.1] (1,1.75) arc (-180:-90-25:1) -- (2,1.75) -- cycle;
\node at (2,1.75) (a) {};
\path (a) ++(-90-25:1.6) node (c) {\scriptsize{\ding{194}}};
\node at (0.75,0.75)  {\scriptsize{$\mathcal{S}_3 = \mathcal{S}_3^+$}};
\node at (2.08,0.75)  {\scriptsize{$\mathcal{S}_3 = 0$}};

\fill[opacity=0.4] (1+4.5,1.75) arc (180:180-25:1) -- (2+4.5,1.75) -- cycle;
\fill[opacity=0.1] (3+4.5,1.75) arc (0:180-25:1) -- (2+4.5,1.75) -- cycle;
\fill[white] (3+4.5,1.75) arc (0:90-25:1) -- (2+4.5,1.75) -- cycle;
\draw[line width=0.50mm,dashed] (2+4.5,1.75) -- ++(-25:1.40);
\draw[line width=0.50mm,dashed] (2+4.5,1.75) -- ++(-180-25:1.40);
\draw[line width=0.50mm] (2+4.5,1.75) -- ++(-25+90:1.40);
\draw[->] (2+4.5,0.25) -- (2+4.5,3.25);
\draw[->] (0.5+4.5,1.75) -- (3.5+4.5,1.75);
\draw[decoration = {zigzag,segment length = 1mm, amplitude = 0.25mm},decorate, line width=0.5mm,black] (2+4.5,1.75)--(0.65+4.5,1.75);
\node at (2+4.5,1.75) (a) {};
\path (a) ++(90-25:1.6) node (c) {\scriptsize{\ding{195}}};
\node at (3.25,0.75+4)  {\scriptsize{$\mathcal{S}_1 = 0$}};
\node at (1.425,0.55+4)  {\scriptsize{$\mathcal{S}_1 = \mathcal{S}_1^+$}};

\draw[line width=0.50mm,dashed] (2+4.5,1.75+4) -- ++(25:1.40);
\draw[line width=0.50mm,dashed] (2+4.5,1.75+4) -- ++(-180+25:1.40);
\draw[line width=0.50mm] (2+4.5,1.75+4) -- ++(25+90:1.40);
\draw[->] (2+4.5,0.25+4) -- (2+4.5,3.25+4);
\draw[->] (0.5+4.5,1.75+4) -- (3.5+4.5,1.75+4);
\draw[decoration = {zigzag,segment length = 1mm, amplitude = 0.25mm},decorate, line width=0.5mm,black] (2+4.5,1.75+4)--(0.65+4.5,1.75+4);
\fill[opacity=0.1] (1+4.5,1.75+4) arc (180:90+25:1) -- (2+4.5,1.75+4) -- cycle;
\node at (2+4.5,1.75+4) (a) {};
\path (a) ++(90+25:1.6) node (c) {\scriptsize{\ding{193}}};
\node at (0.75+4.5,3.5-0.75+4)  {\scriptsize{$\mathcal{S}_2 = \mathcal{S}_2^+$}};
\node at (2.08+4.5,3.5-0.75+4)  {\scriptsize{$\mathcal{S}_2 = 0$}};

\fill[opacity=0.4] (1,1.75+4) arc (180:180+25:1) -- (2,1.75+4) -- cycle;
\fill[opacity=0.1] (3,1.75+4) arc (0:-180+25:1) -- (2,1.75+4) -- cycle;
\fill[white] (3,1.75+4) arc (0:-90+25:1) -- (2,1.75+4) -- cycle;
\draw[line width=0.50mm,dashed] (2,1.75+4) -- ++(25:1.40);
\draw[line width=0.50mm,dashed] (2,1.75+4) -- ++(-180+25:1.40);
\draw[line width=0.50mm] (2,1.75+4) -- ++(25-90:1.40);
\draw[->] (2,0.25+4) -- (2,3.25+4) node [left] {\scriptsize{$\mathrm{Im}(s)$}};
\draw[->] (0.5,1.75+4) -- (3.5,1.75+4) node [below] {\scriptsize{$\mathrm{Re}(s)$}};;
\draw[decoration = {zigzag,segment length = 1mm, amplitude = 0.25mm},decorate, line width=0.5mm,black] (2,1.75+4)--(0.65,1.75+4);
\node at (2,1.75+4) (a) {};
\path (a) ++(-90+25:1.6) node (c) {\scriptsize{\ding{192}}};
\node at (8.5-0.75,3.5-0.75)  {\scriptsize{$\mathcal{S}_4 = 0$}};
\node at (8.5-2.75,3.5-0.55)  {\scriptsize{$\mathcal{S}_4 = \mathcal{S}_4^+$}};

\fill[opacity=0.4] (1+9,1.75) arc (-180:-180+25:1) -- (2+9,1.75) -- cycle;
\fill[opacity=0.4] (1+9,1.75) arc (180:180-25:1) -- (2+9,1.75) -- cycle;
\fill[opacity=0.1] (3+9,1.75) arc (0:180-25:1) -- (2+9,1.75) -- cycle;
\fill[opacity=0.1] (3+9,1.75) arc (0:-180+25:1) -- (2+9,1.75) -- cycle;
\fill[white] (3+9,1.75) arc (0:90-25:1) -- (2+9,1.75) -- cycle;
\fill[white] (3+9,1.75) arc (0:-90+25:1) -- (2+9,1.75) -- cycle;
\draw[line width=0.50mm,dashed] (2+9,1.75) -- ++(180-25:1.40);
\draw[line width=0.50mm,dashed] (2+9,1.75) -- ++(180+25:1.40);
\draw[line width=0.50mm,dashed] (2+9,1.75) -- ++(180-25:1.40);
\draw[line width=0.50mm] (2+9,1.75) -- ++(90+25:1.40);
\draw[line width=0.50mm] (2+9,1.75) -- ++(90-25:1.40);
\draw[line width=0.50mm] (2+9,1.75) -- ++(-90+25:1.40);
\draw[line width=0.50mm] (2+9,1.75) -- ++(-90-25:1.40);
\draw[->] (2+9,0.25) -- (2+9,3.25);
\draw[->] (0.5+9,1.75) -- (3.5+9,1.75);
\draw[decoration = {zigzag,segment length = 1mm, amplitude = 0.25mm},decorate, line width=0.5mm,black] (2+9,1.75)--(0.65+9,1.75);
\node at (2+9,1.75) (a) {};
\path (a) ++(-90+25:1.6) node (c) {\scriptsize{\ding{192}}};
\path (a) ++(90+25:1.6) node (c) {\scriptsize{\ding{193}}};
\path (a) ++(90-25:1.6) node (c) {\scriptsize{\ding{195}}};
\path (a) ++(-90-25:1.6) node (c) {\scriptsize{\ding{194}}};
\draw[->] (2+9+0.75,1.75) arc (0:180-25:0.75); 
\draw[->] (2+9+0.75,1.75) arc (0:-180+25:0.75);
\node at (2+9+0.675,1.75) [above right] {\scriptsize{Valid}};

\node at (2,0) [below] {\scriptsize{$\chi_3$}};
\node at (6.5,0) [below] {\scriptsize{$\chi_4$}};
\node at (2,7.5) [above] {\scriptsize{$\chi_1$}};
\node at (6.5,7.5) [above] {\scriptsize{$\chi_2$}};
\node at (11,0) [below] {\scriptsize{Composite Behaviour}};
\end{tikzpicture}
}

\caption{Behaviour of the singulants for Type B solutions in the complex $s$-plane. We illustrate the behaviour of the four exponential contributions (associated with the four singulants) individually, and also the composite behaviour when each is taken into account. The region in which the series expansion (\ref{1:seriesinf}) provides a valid asymptotic approximation for $u(s,\eps)$ and $v(s,\eps)$ is illustrated in the composite expression. In each figure, the change in Stokes multiplier $\mathcal{S}$ across Stokes lines is illustrated. We recall that the jump across Stokes lines is specified in (\ref{4:Sjump}). Hence, as $\mathcal{S}_{2,3}$ must be zero on the real axis, their behaviour is completely specified. However, this is not true of $\mathcal{S}_{1,4}$, which leads to two free parameters being present in the problem (specifically, the value of these Stokes multipliers on the real axis). Figure (a) illustrates the behaviour for general, non-zero choices of $\mathcal{S}_{1,4}$ on the real axis. Figure (b) illustrates a special case when $\mathcal{S}_{1,4}$ are chosen to be zero on the real axis, giving a larger region of validity for the asymptotic solution.}\label{Stokes_dPIab2}
\end{figure}

Consequently, we determine that the exponential contributions active in each region of the plane are those shown in Figure \ref{Stokes_dPIab2}(a). We see that, assuming that $\mathcal{S}_{1,4}$ is nonzero on the real axis, the region in which the asymptotic expansion associated with (\ref{1:earlyordersb}) is valid is a sector between the anti-Stokes lines associated with $\chi_{1,2}$ and $\chi_{3,4}$, in which the exponentials associated with $\chi_2$ and $\chi_3$ are inactive, and those associated with $\chi_1$ and $\chi_4$ are exponentially small. The angle between the edge of the sector and the real axis (denoted by $\theta$ in Figure \ref{Stokes_dPIab2}) is given by
\begin{equation*}
\theta = \tan^{-1}\left(\frac{\log(2+\sqrt{3})}{\pi}\right) \approx 22.74^{\circ}.
\end{equation*}

However, we note that it is possible to choose $\mathcal{S}_{1,4} = 0$ on the real axis. Figure \ref{Stokes_dPIab2}(b) illustrates this behaviour. In this case, the region in which the asymptotic expansion is valid is substantially larger than the general case shown in Figure \ref{Stokes_dPIab2}(a); there is no exponential contribution associated with $\chi_1$ or $\chi_4$ to become large as the anti-Stokes lines are crossed. Instead, the angle $\theta$ is now given by
\begin{equation*}
\theta = \pi - \tan^{-1}\left(\frac{\log(2+\sqrt{3})}{\pi}\right) \approx 157.26^{\circ}.
\end{equation*}

\subsection{Stokes smoothing}\label{1.4:Smoothing}

Using a nearly identical analysis to Section \ref{2.4:Smoothing}, we are able to show that for Type B solutions, the quantity switched across each Stokes line takes the form
\begin{equation}\label{1:rng0}
R_N \sim \mathcal{S}  U \e^{-\chi/\eps},
\end{equation}
where $R_N$ is again the exponential contribution to the solution, and $\mathcal{S} $ is the Stokes switching parameter. Using matched asymptotic expansions, we find that
\begin{equation*}
[\mathcal{S} ]^+_- \sim \frac{6\pi}{\chi'\sinh(-\chi')}.
\end{equation*}
We note that
\begin{align*}
\chi_{1,2}'\sinh(-\chi_{1,2}') = \sqrt{3}\log(-2-\sqrt{3}),\qquad
\chi_{3,4}'\sinh(-\chi_{3,4}') = -\sqrt{3}\log(-2+\sqrt{3}).
\end{align*}
Hence, from (\ref{1:rng0}), the full exponential behaviour takes the form
\begin{align*}
R_N \sim \mathcal{S} _1 &\Lambda_1  (-2-\sqrt{3})^{-s/\eps} + \mathcal{S} _2 \Lambda_2 (-2-\sqrt{3})^{s/\eps} \\&+ \mathcal{S} _3\Lambda_3 (-2+\sqrt{3})^{-s/\eps} + \mathcal{S} _4\Lambda_4  (-2+\sqrt{3})^{s/\eps}, 
\end{align*}
where $\mathcal{S}_i$ switches rapidly across the relevant Stokes line such that the difference is given by
\begin{align}\label{4:Sjump}
[\mathcal{S} _{1,2}]^+_- \sim \frac{2\pi\sqrt{3}}{\log(-2-\sqrt{3})},\qquad
[\mathcal{S} _{3,4}]^+_- \sim \frac{-2\pi\sqrt{3}}{\log(-2+\sqrt{3})}.
\end{align}
We also note for completion that $\mathcal{S} _2$ and $\mathcal{S} _3$ must equal zero on one side of the Stokes line, and hence are uniquely determined, while this condition is not required for $\mathcal{S} _1$ and $\mathcal{S} _4$. However, if both $\mathcal{S} _1^+$ and $\mathcal{S} _4^+$ are zero, we obtain the special Stokes structure illustrated in Figure \ref{Stokes_dPIab2}(b).

Consequently, the optimally-truncated asymptotic series for $u(s,\eps)$, and hence $v(s,\eps)$, is shown to take the form
\begin{align}\nonumber
u(s,\eps) \sim \sum_{m=0}^{N_{\mathrm{opt}}} \eps^{m/2} u_m(s) + \mathcal{S} _1 &\Lambda_1  (-2-\sqrt{3})^{-s/\eps} + \mathcal{S} _2 \Lambda_2 (-2-\sqrt{3})^{s/\eps} \\&+ \mathcal{S} _3\Lambda_3 (-2+\sqrt{3})^{-s/\eps} + \mathcal{S} _4\Lambda_4  (-2+\sqrt{3})^{s/\eps}\label{4:asymp2app}
\end{align}
where the leading orders of $u_n$ are give in (\ref{1:earlyordersb}), and the late-order behaviour is given in (\ref{4:LOTcalc}). This expression is an accurate asymptotic approximation up to exponentially-small terms. Unlike the Type A solution, however, this approximation of the Type B solution is only valid within a wedge of the complex plane illustrated in Figure \ref{Stokes_dPIab2}. We also recall from Section \ref{1.3:Structure} that $\mathcal{S}_2$ and $\mathcal{S}_3$ are uniquely determined, while $\mathcal{S}_1$ and $\mathcal{S}_4$ contain a Stokes-switching parameter that may be freely specified; this behaviour is not present in the asymptotic approximation of Type A solutions. 

\section{Conclusions}

In this paper, we used methods of exponential asymptotics to compute and investigate two types of asymptotic solutions to the first discrete Painlev\'e equation, associated with the leading order behaviours described in (\ref{1:earlyordersa}) and (\ref{1:earlyordersb}). We identified the Stokes structure of the solution for each type, and used this information to deduce the sectors displaying asymptotically pole-free behaviour in the far-field limit. These asymptotic solutions were given as the sum of a truncated asymptotic power series and an exponentially-subdominant remainder term, shown in (\ref{4:asymp1app})-(\ref{4:asymp1app2}) for Type A solutions, and (\ref{4:asymp2app}) for Type B solutions.

In Section \ref{CASEA}, we showed that the asymptotic solution associated with (\ref{1:earlyordersa}) is a valid approximation in the far field of the entire complex plane, aside from a branch-cut along the negative real axis, and hence the solution is asymptotically pole-free. This solution has no exponentially-small behaviour on the positive real axis, although there is exponentially-small behaviour present in the left-half complex plane. This implies that this asymptotic behaviour is uniquely defined, with no free parameter present in the solution.

The asymptotic solution associated with (\ref{1:earlyordersb}), discussed in Section \ref{CASEB}, is generally asymptotically pole-free within a narrow wedge in the complex plane containing the positive real axis, seen in Figure \ref{Stokes_dPIab2}(a). Unlike the first asymptotic solution, however, there are two free parameters in the asymptotic approximation, associated with Stokes switching. If these parameters are chosen appropriately, we discover asymptotic solutions with larger ranges of validity, such as that shown in Figure \ref{Stokes_dPIab2}(b). Outside of the asymptotically pole-free regions, the asymptotic series solution is no longer valid, and we instead see the elliptic function behaviour described by \cite{Joshi2}. In fact, the oscillatory behaviour that is present along anti-Stokes lines is a degenerate form of this behaviour.

We note with interest that the two classes of asymptotically pole-free solutions to dP$_{\mathrm{I}}$, Types A and B, demonstrate power series-type behaviour in the far field, however they differ in that Type A solutions are uniquely specified with no free parameters, while Type B solutions contain free parameters associated with the Stokes multipliers in the solution. Similar asymptotic behaviour is seen in the tritronqu\'ee and tronqu\'ee solutions to P$_{\mathrm{I}}$, the former of which is uniquely specified, while the latter contains free parameters associated with Stokes switching. Consequently, the two classes of asymptotic behaviour described in this study may be regarded as corresponding to tritronqu\'ee and tronqu\'ee solutions of P$_{\mathrm{I}}$ respectively.

The asymptotic solutions to dP$_{\mathrm{I}}$ also provide insight into the behaviour of solution hierarchies to the equations described in (\ref{0:P4dP1}) and (\ref{0:pndPI}). The first of these describes solution hierarchies to the fourth Painlev\'e equation. For example, Type B solutions (\ref{1:earlyordersb}) correspond to solution families which grow as $w_n(z) \sim \pm\sqrt{-n/3} - z/3 + \mathcal{O}(n^{-1/2})$ in the large-$n$ limit, with the parameters described in (\ref{0:P4par}) also growing appropriately. These solution hierarchies are of particular mathematical interest due to their power series dependence on $n$ in the asymptotic limit, which is particular to the asymptotic classes of dP$_{\mathrm{I}}$ solutions described in the present study.

Finally, we emphasize that this methodology may be applied to a wide range of difference equations, as it does not require the problem to be integrable, nor does it require that solution be expressed in terms of an integral equation. This is demonstrated in Appendix \ref{A3}, where we outline how this procedure could be applied to determine the Stokes switching behaviour of a non-integrable difference equation. Consequently, although we applied the method to determine information about the exponentially-small components of the solution to dP$_{\mathrm{I}}$, it could easily be adapted to analyze asymptotic solutions of broad classes of difference equations. 

One particular case of interest is the behaviour of orthonormal polynomials satisfying (\ref{0:pnw}) for a variety of different weight functions, such as the semi-classical weights described in \cite{Magnus2}. In general, these polynomial families may be computed using non-integrable difference equations. However, as there is no aspect of the methodology which depends on the integrability of the equation, these difference equations are still amenable to the asymptotic methods described in the present study. 


\section{Data Accessibility}
We have no supporting data aside from that contained within the text.

\section{Competing Interests}
We have no competing interests.

\section{Authors' Contributions}
NJ and CJL collaborated on the mathematical analysis. CJL drafted the manuscript. All authors gave final approval for publication.

\section{Acknowledgments}
The authors would like to thank Prof. C. J. Howls and Prof. Y. Takei for discussion and suggestions regarding this study. The authors would also like to thank the referees for their helpful comments and suggestions.

\section{Funding Statement}
NJ and CJL were supported by Australian Laureate Fellowship Grant \#FL120100094 from the Australian Research Council.

\appendix

\section{Calculating the prefactor constants}\label{1:2Inner}

\subsection{Type A}\label{A_typeA}

To determine the value of the constants $\Lambda_{i}$, we return to the late-order expression given in (\ref{2:LOT}). We see that these expressions may be rewritten as
\begin{equation}\label{A:LOT1}
\Lambda_1 - \Lambda_2(-1)^{m/2-1/2} \sim \frac{u_m (\mathrm{i}\pi s/2)^{m/2-1/2}}{\Gamma(m/2-1/2)},\qquad
\Lambda_3 - \Lambda_4(-1)^{m/2-1/2} \sim \frac{v_m (\mathrm{i}\pi s/2)^{m/2-1/2}}{\Gamma(m/2-1/2)}.
\end{equation}

By adding (and subtracting) successive terms appropriately in the limit $m \rightarrow \infty$, we may easily determine $\Lambda_1$ and $\Lambda_2$ by considering sufficiently high values of $u_m$. For example, adding and subtracting appropriate even terms gives
\begin{subequations}
\begin{align}\label{A:ajlim1b}
2\Lambda_1 &= \lim_{m\rightarrow \infty} \left[\frac{u_{2m} (\mathrm{i}\pi s/2)^{m-1/2}}{\Gamma(m-1/2)}-\frac{u_{2m+2} (\mathrm{i}\pi s/2)^{m+1/2}}{\Gamma(m+1/2)}\right],\\
2\mathrm{i}\Lambda_2 &= \lim_{m\rightarrow \infty} \left[ \frac{u_{2m} (\mathrm{i}\pi s/2)^{m-1/2}}{\Gamma(m-1/2)}+\frac{u_{2m+2} (\mathrm{i}\pi s/2)^{m+1/2}}{\Gamma(m+1/2)}\right].\label{A:ajlim2b}
\end{align}
\end{subequations}
Similar results may be obtained for $\Lambda_3$ and $\Lambda_4$ by considering the behaviour of $v_n$. 

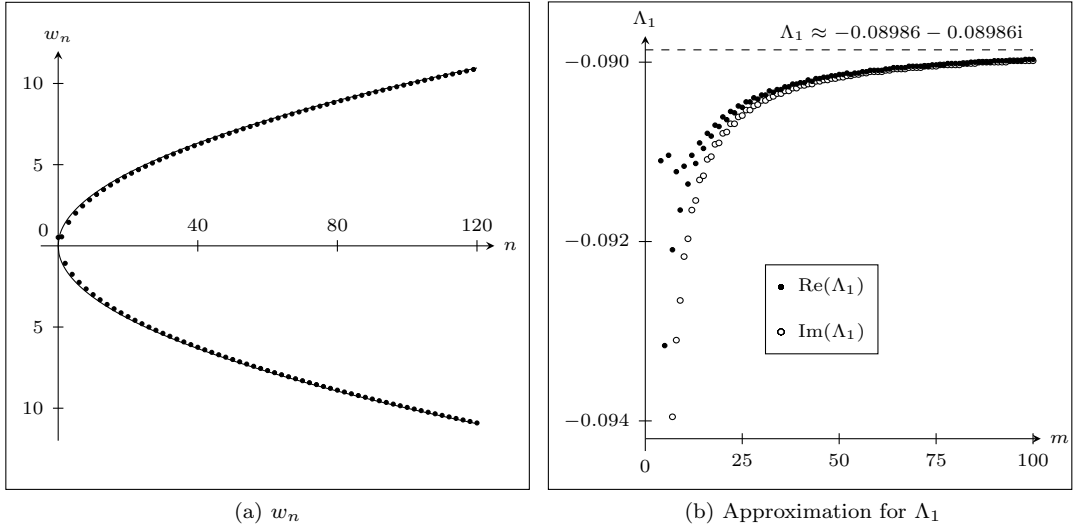
\begin{figure}
\begin{center}

\subfloat[$w_n$]{
\centering
\begin{tikzpicture} 
[xscale=0.054*0.85,>=stealth,yscale=0.2533*0.85]

\draw(-15,15) -- (135,15) -- (135,-15) -- (-15,-15) -- cycle;

\draw[->] (-5,0) -- (125,0) node[right] {\scriptsize{$n$}};
\draw [->] (0,-12) -- (0,12) node[above] {\scriptsize{$w_n$}};

\draw[] plot[smooth] file {sqrtan.txt};
\draw[] plot[smooth] file {msqrtan.txt};
\draw[] plot[only marks, mark=*,mark options={xscale=(1/0.054)*0.45,yscale=(1/0.2533)*0.45}] file {a1b1q00q11.txt};

\draw (40,-0.35) -- (40,0.35) node[above] {\scriptsize{$40$}};
\draw (80,-0.35) -- (80,0.35) node[above] {\scriptsize{$80$}};
\draw (120,-0.35) -- (120,0.35) node[above] {\scriptsize{$120$}};
\node at (0,0) [above left] {\scriptsize{$0$}};

\draw(-1.5,5) node[left] {\scriptsize{$5$}} -- (1.5,5);
\draw(-1.5,10) node[left] {\scriptsize{$10$}} -- (1.5,10);
\draw(-1.5,-5) node[left] {\scriptsize{$5$}} -- (1.5,-5);
\draw(-1.5,-10) node[left] {\scriptsize{$10$}} -- (1.5,-10);

\end{tikzpicture}
}
\subfloat[Approximation for $\Lambda_1$]{
\centering
\begin{tikzpicture} 
[xscale=0.06* 0.85,>=stealth,yscale=1400* 0.85]

\draw(-25,-0.09475) -- (110,-0.09475) -- (110,-0.09475+0.00542857)-- (-25,-0.09475+0.00542857) -- cycle;

\draw[->] (0 ,-0.0942) -- (102, -0.0942) node[right] {\scriptsize{$m$}};
\draw [<-] (0,-0.0897) node[above] {\scriptsize{$\Lambda_1$}} -- (0,-0.0942);

\draw[black] plot[only marks, mark=*,mark options={fill=white,xscale= 16.667*0.65,yscale=0.0007*0.65}] file {aratimag.txt};
\draw[black] plot[only marks, mark=*,mark options={xscale= 16.667*0.5,yscale=0.0007*0.5}] file {aratreal.txt};
\draw[dashed,black] (0,-0.08986) -- (100,-0.08986) node[above left] {\scriptsize{$\Lambda_1 \approx -0.08986 -0.08986\mathrm{i}$}};

\draw(-3.33/2, -0.094) node[left] {\scriptsize{$-0.094 $}} -- (3.33/2, -0.094);
\draw(-3.33/2, -0.092) node[left] {\scriptsize{$-0.092$}} -- (3.33/2, -0.092);
\draw(-3.33/2, -0.090) node[left] {\scriptsize{$-0.090$}} -- (3.33/2, -0.090);

\draw(0,-0.0942) -- (0,-0.0942-0.000142) node[below] {\scriptsize{$0$}};
\draw(50,-0.0942+0.000142/2) -- (50,-0.0942-0.000142/2) node[below] {\scriptsize{$50$}};
\draw(25,-0.0942+0.000142/2) -- (25,-0.0942-0.000142/2) node[below] {\scriptsize{$25$}};
\draw(75,-0.0942+0.000142/2) -- (75,-0.0942-0.000142/2) node[below] {\scriptsize{$75$}};
\draw(100,-0.0942+0.000142/2) -- (100,-0.0942-0.000142/2) node[below] {\scriptsize{$100$}};

\draw[black] plot[only marks, mark=*,mark options={fill=white,xscale= 16.667*0.65,yscale=0.0007*0.65}] (25+10,-0.093);
\node at (27+10,-0.093) [right] {\scriptsize{$\mathrm{Im}(\Lambda_1)$}};
\draw[black] plot[only marks, mark=*,mark options={xscale=16.667*0.5,yscale=0.0007*0.5}] (25+10,-0.0925);
\node at (27+10,-0.0925) [right] {\scriptsize{$\mathrm{Re}(\Lambda_1)$}};

\draw(21+10,-0.09225) -- (50+10,-0.09225) -- (50+10,-0.09325) -- (21+10,-0.09325) -- cycle;

\end{tikzpicture}
}

\end{center}
\caption{Figure (a) illustrates the behaviour of the solution to dP$_\mathrm{I}$ with $\alpha = -1$,  $\beta = 1$ and $\gamma = 0$. The boundary conditions are chosen such that $w_0 = 0$ and $w_1$ = 1. The values of $w_n$ are represented as black circles, and the asymptotic solution $w_n = \pm\sqrt{-\alpha n}$, associated with Type A solutions, is represented as a solid curve. It is apparent that the behaviour of the discrete equation tends to the asymptotic expression in question for large $n$. Figure (b) illustrates the approximation for $\Lambda_1$ obtained by computing the expression given in (\ref{A:ajlim1b}). As $m$ increases, the approximation tends to the dashed curve.}
\label{Afig}
\end{figure}

As a specific example, we consider the case when $\alpha = -1$, $\beta = 1$ and $\gamma = 0$. A numerical study similar to that performed by Joshi and Kitaev \cite{Joshi3}, shown in Figure \ref{Afig}(a), indicates that this behaviour tends to the appropriate asymptotic expression with the boundary conditions $w_0 = 0.52040003$ and $w_1 = 0.55549107$. We select the case when $u_0 = \sqrt{s}$, giving $u_1 = 0$ and $u_2 = -1/2\sqrt{s}$. We may now evaluate (\ref{1:orderna})-(\ref{1:ordernb}) in order to determine progressively higher terms in the series. In Figure \ref{Afig}(b), we apply (\ref{A:ajlim1b}) to approximate $\Lambda_1$ for various values of $m$, with the dashed line indicating the quantity that $\Lambda_1$ converges to for sufficiently large $m$. Using the general relation in (\ref{1:orderna})-(\ref{1:ordernb}), we calculate behaviour of $u_m$ up to $m = 251$, we find

\begin{equation*}
\Lambda_1 \approx -0.08986 - 0.08986\mathrm{i}, \qquad \Lambda_2 \approx - 0.08986 + 0.08986\mathrm{i}.
\end{equation*}
As $\gamma = 0$, we have $u_m = -v_m$, we also conclude that $\Lambda_3 = -\Lambda_1$ and $\Lambda_4 = -\Lambda_2$. This limit may be computed for different values of $\alpha$ and $\beta$, although convergence can be somewhat slow for some choices of parameters. Finally, if we permit $\gamma$ to take nonzero values, we must consider the late-term behaviour of $u_m$ and $v_m$ independently, and solve for the four values of $\Lambda_i$ separately.

\subsection{Type B}\label{A_typeB}
When considering the more complicated form of the late-order terms for Type B solutions given in (\ref{4:LOTcalc}), we must solve for four different values of $\Lambda_i$. In this case, we write $\chi_1 = |\chi_1|\e^{\mathrm{i}\theta}$, and note the late-order terms may be rewritten as
\begin{equation}\label{A:Lambda}
\Lambda_1 - (-1)^{m/2}\Lambda_2 + \e^{\mathrm{i}\theta(m-1)}\Lambda_3 - (-1)^{m/2}\e^{\mathrm{i}\theta(m-1)}\Lambda_4 \sim \frac{u_m (\log(-2-\sqrt{3})s)^{m/2-1/2}}{\Gamma(m/2-1/2)}.
\end{equation}
While this expression is not as simple as (\ref{A:LOT1}), we may still use it to determine the values of $\Lambda_i$ by taking four distinct values of $m$ and solving the resultant system of equations. As these values of $m$ become large, the computed values for $\Lambda_i$ will converge to the appropriate values. 

\begin{figure}
\begin{center}

\subfloat[Approximation for $\Lambda_1$]{
\centering
\begin{tikzpicture} 
[xscale=0.06* 0.85,>=stealth,yscale=95* 0.85]

\draw(-20,-0.01) -- (115,-0.01) -- (115,0.07)-- (-20,0.07) -- cycle;

\draw[->] (0 ,0) -- (105, 0) node[right] {\scriptsize{$m$}};
\draw [<-] (0,0.065) node[above] {\scriptsize{$\Lambda_1$}} -- (0,-0.0025);

\draw[dashed,black] (0,0.04666) -- (105,0.04666) node[above left] {\scriptsize{$\mathrm{Re}(\Lambda_1) \approx 0.04666$}};
\draw[dashed,black] (0,0.02669) -- (105,0.02669) node[above left] {\scriptsize{$\mathrm{Im}(\Lambda_1) \approx 0.02669$}};

\draw[black] plot[only marks, mark=*,mark options={fill=white,xscale= 16.667*0.6,yscale=0.0103*0.6}] file {a3b1L1_imag.txt};
\draw[black] plot[only marks, mark=*,mark options={xscale=16.667*0.65,yscale=0.0103*0.65}] file {a3b1L1_real.txt};

\draw(-3.33/2, 0) node[left] {\scriptsize{$0 $}} -- (3.33/2, 0);
\draw(-3.33/2, 0.02) node[left] {\scriptsize{$0.02$}} -- (3.33/2, 0.02);
\draw(-3.33/2, 0.04) node[left] {\scriptsize{$0.04$}} -- (3.33/2, 0.04);
\draw(-3.33/2, 0.06) node[left] {\scriptsize{$0.06$}} -- (3.33/2, 0.06);

\draw(0,0.00125) -- (0,-0.0025) node[below] {\scriptsize{$0$}};
\draw(25,0.00125) -- (25,-0.00125) node[below] {\scriptsize{$25$}};
\draw(50,0.00125) -- (50,-0.00125) node[below] {\scriptsize{$50$}};
\draw(75,0.00125) -- (75,-0.00125) node[below] {\scriptsize{$75$}};
\draw(100,0.00125) -- (100,-0.00125) node[below] {\scriptsize{$100$}};

\draw[black] plot[only marks, mark=*,mark options={fill=white,xscale= 16.667*0.65,yscale=0.0103*0.65}] (75,0.01);
\node at (75,0.01) [right] {\scriptsize{$\mathrm{Im}(\Lambda_1)$}};
\draw[black] plot[only marks, mark=*,mark options={xscale=16.667*0.5,yscale=0.0103*0.5}] (75,0.018);
\node at (75,0.018) [right] {\scriptsize{$\mathrm{Re}(\Lambda_1)$}};

\draw(71,0.006) -- (98,0.006) -- (98,0.022) -- (71,0.022) -- cycle;

\end{tikzpicture}
}
\subfloat[Approximation for $\Lambda_4$]{
\centering
\begin{tikzpicture} 
[xscale=0.06* 0.85,>=stealth,yscale=95* 0.85]

\draw(-20,-0.01+0.06) -- (115,-0.01+0.06) -- (115,0.07+0.06)-- (-20,0.07+0.06) -- cycle;

\draw[->] (0 ,0+0.06) -- (105, 0+0.06) node[right] {\scriptsize{$m$}};
\draw [<-] (0,0.065+0.06) node[above] {\scriptsize{$\Lambda_4$}} -- (0,-0.0025+0.06);

\draw[dashed,black] (0,0.11212) -- (105,0.11212) node[above left] {\scriptsize{$\mathrm{Re}(\Lambda_4) \approx 0.11212$}};
\draw[dashed,black] (0,0.06491) -- (105,0.06491) node[above left] {\scriptsize{$\mathrm{Im}(\Lambda_4) \approx 0.06491$}};

\draw[black] plot[only marks, mark=*,mark options={fill=white,xscale= 16.667*0.6,yscale=0.0103*0.6}] file {a3b3L4_imag.txt};
\draw[black] plot[only marks, mark=*,mark options={xscale=16.667*0.65,yscale=0.0103*0.65}] file {a3b3L4_real.txt};

\draw(-3.33/2, 0.08) node[left] {\scriptsize{$0.08$}} -- (3.33/2, 0.08);
\draw(-3.33/2, 0.10) node[left] {\scriptsize{$0.10$}} -- (3.33/2, 0.10);
\draw(-3.33/2, 0.12) node[left] {\scriptsize{$0.12$}} -- (3.33/2, 0.12);
\draw(-3.33/2, 0.06) node[left] {\scriptsize{$0.06$}} -- (3.33/2, 0.06);

\draw(0,0.00125+0.06) -- (0,-0.0025+0.06) node[below] {\scriptsize{$0$}};
\draw(25,0.00125+0.06) -- (25,-0.00125+0.06) node[below] {\scriptsize{$25$}};
\draw(50,0.00125+0.06) -- (50,-0.00125+0.06) node[below] {\scriptsize{$50$}};
\draw(75,0.00125+0.06) -- (75,-0.00125+0.06) node[below] {\scriptsize{$75$}};
\draw(100,0.00125+0.06) -- (100,-0.00125+0.06) node[below] {\scriptsize{$100$}};

\draw[black] plot[only marks, mark=*,mark options={fill=white,xscale= 16.667*0.65,yscale=0.0103*0.65}] (75,0.01+0.06+0.02);
\node at (75,0.01+0.06+0.02) [right] {\scriptsize{$\mathrm{Im}(\Lambda_4)$}};
\draw[black] plot[only marks, mark=*,mark options={xscale=16.667*0.5,yscale=0.0103*0.5}] (75,0.018+0.06+0.02);
\node at (75,0.018+0.06+0.02) [right] {\scriptsize{$\mathrm{Re}(\Lambda_4)$}};

\draw(71,0.006+0.06+0.02) -- (98,0.006+0.06+0.02) -- (98,0.022+0.06+0.02) -- (71,0.022+0.06+0.02) -- cycle;

\end{tikzpicture}
}

\end{center}
\caption{This figure illustrates the computed approximations associated with Type B solutions having $\alpha = 3$, $\beta = 1$ and $\gamma = 0$ for (a) $\Lambda_1$ and (b) $\Lambda_4$, obtained by solving the system generated from (\ref{A:Lambda}) using four consecutive even values of $m$, beginning with the value indicated on the horizontal axis. As $m$ increases, the approximations tend to the dashed curve.}
\label{Afig2}
\end{figure}
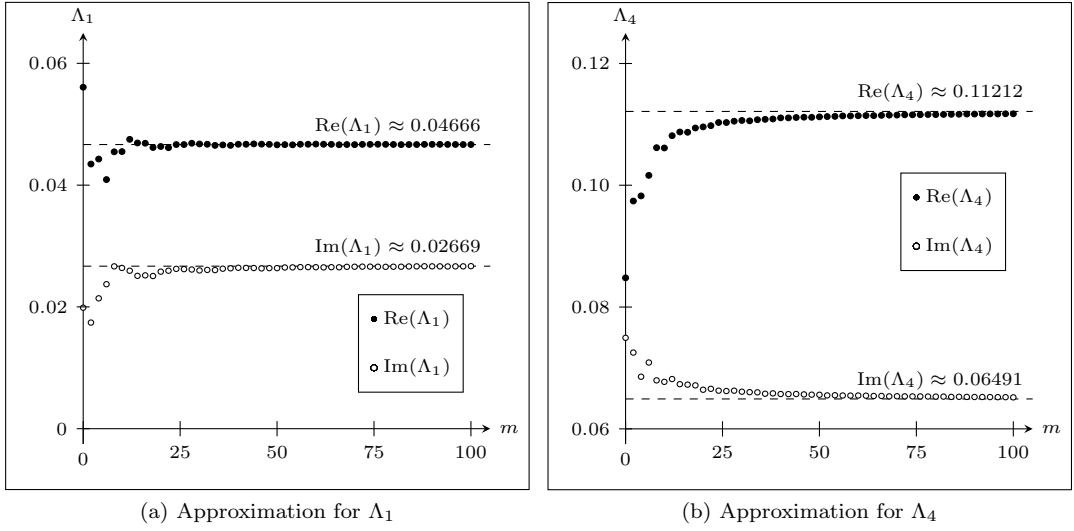

We consider a numerical example, choosing $\alpha = 3$, $\beta = 1$ and $\gamma = 0$ (ensuring that $u_m = v_m$, and hence eliminating the need to consider $v_m$ separately). We begin with the leading-order solution $u_0 = \sqrt{s}$ and compute the behaviour of $u_m$ up to $m = 500$ and use four consecutive even values of $m$ in (\ref{A:Lambda}) to generate the system of equations which is subsequently solved for $\Lambda_i$. The computed values of $\Lambda_1$ and $\Lambda_4$ are shown in Figure \ref{Afig2} for a range of values of $m$, and we see that they are converging to a constant value. By taking a system of equations using sufficiently large computed values of $u_m$, we find that
\begin{align*}
\Lambda_1& \approx 0.04666 + 0.02669\mathrm{i},\quad &\Lambda_2& \approx 0.11212 - 0.06491\mathrm{i},\\
\Lambda_3 & \approx 0.04666 - 0.02669\mathrm{i},\quad &\Lambda_4& \approx 0.11212 + 0.06491\mathrm{i}.
\end{align*}
We have therefore determined the form of the late-order terms $u_m$, and consequently $v_m$. Again, if we considered an example with $\gamma \neq 0$, we would find that $u_m \neq v_m$, and the values of $\Lambda_i$ associated with $v_m$ would need to be computed separately.

\section{Stokes smoothing}\label{A2}

To perform an exponential asymptotic analysis on the remainder terms in (\ref{4:seriesN}), we must ensure that the asymptotic series is truncated optimally. To find the optimal truncation point, we follow the commonly-used strategy described by \cite{Boyd1}, in which the series is truncated at its smallest term (and hence where consecutive terms have the same size). 

By solving 
\begin{equation*}
\left|\frac{\eps^{(N+1)/2} u_{N+1}}{\eps^{N}u_{N}} \right|\sim 1,
\end{equation*}
in the limit $\eps \rightarrow 0$ and $N \rightarrow \infty$, we find that $N \sim 2|\chi|/\eps$. Hence, we will set 
\begin{equation}\label{4:Nopt}
N = \frac{2|\chi|}{\eps} + 2\omega,
\end{equation}
where $0 \leq \omega < 1/2$ is chosen so that $N$ takes integer value; the choice of the two in $2\omega$ is one of algebraic convenience.

Applying this series expression and using (\ref{1:orderna})-(\ref{1:ordernb}) to eliminate terms, recalling that $u_0 = -v_0$, gives
\begin{subequations}
\begin{align}
\nonumber 2 u_0 &S_N  + 2\eps^{1/2}  u_1 S_N  + 2\eps u_2 S_N + 2u_0 \sum_{j=1}^{\lfloor N/2 \rfloor} \frac{\eps^{2j} S_N^{(2j)}}{(2j)!}  \\
&+ 2\eps^{1/2} u_1 \sum_{j=1}^{\lfloor N/2 \rfloor-1} \frac{\eps^{2j} S_N^{(2j)} }{(2j)!} + 2\eps u_2 \sum_{j=1}^{\lfloor N/2 \rfloor-1} \frac{\eps^{2j} S_N^{(2j)} }{(2j)!} = \eps^{1/2}\gamma S_N + 2 \eps^{N/2}  v_0 u_N  + \ldots \label{4:PL1a}\\
\nonumber 2 v_0 &R_N  + 2\eps^{1/2}  v_1 R_N  + 2\eps v_2 R_N + 2v_0 \sum_{j=1}^{\lfloor N/2 \rfloor} \frac{\eps^{2j} R_N^{(2j)}}{(2j)!}  \\
&+ 2\eps^{1/2} v_1 \sum_{j=1}^{\lfloor N/2 \rfloor-1} \frac{\eps^{2j} R_N^{(2j)} }{(2j)!}+ 2\eps v_2 \sum_{j=1}^{\lfloor N/2 \rfloor-1} \frac{\eps^{2j} R_N^{(2j)} }{(2j)!} = \eps^{1/2}\gamma R_N + 2 \eps^{N/2}  u_0 v_N  + \ldots \label{4:PL1b}
\end{align}
\end{subequations}
where the omitted terms will be smaller in the limit $\eps \rightarrow 0$ than those which have been retained. 

Noting that $R_N$ and $S_N$ have decoupled, we will use (\ref{4:PL1b}) to determine the behaviour of $R_N$. We apply a WKB ansatz to the homogenous version of (\ref{4:PL1b}) by setting $R_N = f(s)\e^{g(s)/\eps}$. The solution away from the Stokes line (where the right-hand side not have any effect) is therefore given by the homogenous version of (\ref{4:PL1b}), and therefore  to leading order as $\eps \rightarrow 0$
\begin{equation*}
2v_0 f(s)\e^{-g(s)/\eps} + 2 v_0  \sum_{j=1}^{N/2} \left(\frac{g'(s)}{\eps}\right)^{2j}\frac{\eps^{2j}f(s)}{(2j)!}\e^{-g(s)/\eps} = 0.
\end{equation*}
By comparing this to the singulant equation (\ref{LOT:sing}), we see that we can write $g(s) = \chi$, where $\chi$ represents the value of the singulant under consideration; here, the analysis will be performed in the general case, with the particular choice of $\chi$ made afterwards. Continuing to the next order in $\eps$, we will find that $f(s)$ is given by a constant. Hence, away from the Stokes line, the exponentially-small remainder takes the form $R_N \sim C \e^{-\chi/\eps}$, where $C$ is a constant, which we will set to be the associated value for $\Lambda$.

Knowing that the behaviour will switch rapidly in the neighbourhood of the Stokes line, we set
\begin{equation*}
R_N = \mathcal{S} \Lambda \e^{-\chi/\eps},
\end{equation*}
where $\mathcal{S} $ is the Stokes switching parameter that varies rapidly as the Stokes line is crossed. The multiplier $\Lambda$ is chosen for subsequent algebraic convenience. Applying this ansatz to (\ref{4:PL1a}) gives, after some cancellation,
\begin{equation*}
2 \eps  \mathcal{S} ' \Lambda \sum_{j=1}^{\lfloor N/2 \rfloor} \frac{(-\chi')^{2j-1}}{(2j-1)!} = 2\eps^{N/2} \e^{\chi/\eps} u_N.
\end{equation*}
As $N \rightarrow \infty$, the series may be taken to be infinite, introducing only exponential error. However, we have already discarded terms containing algebraic error, so this is not a concern. We therefore write
\begin{equation*}
  \mathcal{S} ' \Lambda \sinh{(-\chi')} \sim \eps^{N/2-1} \e^{\chi/\eps} u_N.
\end{equation*}
Noting the form of $u_N$ given in (\ref{1:ansatz}), we write 
\begin{equation}\label{4:PL1d}
\diff{\mathcal{S} }{s} \sim \eps^{N/2-1}\e^{\chi/\eps}\frac{\Gamma(N/2+k)}{\sinh(-\chi') \chi^{N/2+k}}.
\end{equation}

Noting the form of the optimal truncation (\ref{4:Nopt}), we set $\chi = r\e^{\mathrm{i}\theta}$ and restrict ourselves to curves with $r$ fixed. This gives
\begin{equation*}
\diff{}{s} = -\frac{\mathrm{i}\chi'\e^{-\mathrm{i}\theta}}{r}\diff{}{\theta},
\end{equation*}
while (\ref{4:Nopt}) becomes $N = 2r/\eps + 2\omega$. Hence (\ref{4:PL1d}) becomes
\begin{equation*}
\diff{\mathcal{S} }{\theta} \sim  r \e^{\mathrm{i}\theta} \eps^{r/\eps + \omega -1}\frac{ \Gamma(r/\eps + \omega +k)}{\chi' \sinh(-\chi')  (r\e^{\mathrm{i}\theta})^{r/\eps + \omega +k}}\exp\left(\frac{r\e^{\mathrm{i}\theta}}{\eps}\right).
\end{equation*}
From (\ref{2:singulant}), we note that $\chi'\sinh(-\chi') = \pi/2$ for both $\chi_1$ and $\chi_2$. We also recall that $k = -1/2$. Hence, applying Stirling's formula and simplifying eventually gives
\begin{equation*}
\diff{\mathcal{S} }{\theta} \sim \sqrt{\frac{8r}{\pi}} \exp{\left(\frac{r}{\eps}{(\e^{\mathrm{i}\theta}-1-\mathrm{i}\theta)} -\mathrm{i}\theta(k + \omega - 1)\right)}.
\end{equation*}
This is exponentially small, except in the neighbourhood of $\theta = 0$, which corresponds to the Stokes line (that is, $\chi$ being purely real and positive). To determine the local behaviour, we apply the scaling $\hat{\theta} = \eps^{1/2}\theta$, which gives
\begin{equation*}
\diff{\mathcal{S} }{\theta} \sim \sqrt{\frac{8r}{\pi}} \e^{-r\hat{\theta}^2/2}.
\end{equation*}
Hence, by integration, we find 
\begin{equation}\label{4:Ajump}
\mathcal{S}  \sim \sqrt{\frac{8}{\pi}}\left[\int_{-\infty}^{\hat{\theta}/\sqrt{r}} \e^{-s^2/2}\mathrm{d} s+ C\right],
\end{equation}
where $C$ is an arbitrary constant. Importantly, as the Stokes line is crossed, $\hat{\theta}$ varies rapidly from $-\infty$ to $\infty$. Hence, away from the Stokes line ($\hat{\theta} = 0$), we find that the jump in the Stokes multiplier $A$ as the Stokes line is crossed is given by
\begin{equation*}
[\mathcal{S} ]_-^+ \sim \sqrt{\frac{8}{\pi}}\int_{-\infty}^{\infty} \e^{-s^2/2}\mathrm{d} s = 4 ,
\end{equation*}
and hence that the exponential contribution switched across the Stokes line $R_N$ changes by
\begin{equation*}
\left[R_N\right]_-^+ \sim  4\Lambda\,\e^{-\chi/\eps},
\end{equation*}
as $s$ moves from negative to positive. Recalling that there are exponential contributions from the two singulants given in (\ref{2:singulant}), we find a combined expression
\begin{align}\label{4:RNout}
R_N &\sim  \mathcal{S} _1\Lambda_1\,\e^{-\mathrm{i}\pi s/2\eps} + \mathcal{S} _2\Lambda_2\,\e^{\mathrm{i}\pi s/2\eps},
\end{align}
where $\mathcal{S}_1$ changes rapidly from four to zero when crossing from the third to the fourth quadrant in the smooth manner described by (\ref{4:Ajump}), and $\mathcal{S}_2$ changes from negative four to zero when crossing from the second to the first quadrant in the same way. A similar analysis shows that 
\begin{align}\label{4:RNout2}
R_N &\sim  \mathcal{S} _1\Lambda_3\,\e^{-\mathrm{i}\pi s/2\eps} + \mathcal{S} _2\Lambda_4\,\e^{\mathrm{i}\pi s/2\eps}.
\end{align}

\section{A non-integrable equation}\label{A3}

To demonstrate the utility of this method, we consider a non-integrable variant of dP$_\mathrm{I}$, and briefly outline the steps required to determine the late-order term behaviour, and hence the Stokes structure of the solution. We begin with the equation presented in (\ref{0:dPI}), and adjust the final term, giving
\begin{equation}\label{A:dP}
w_{n+1} + w_n + w_{n-1} = \frac{\alpha n + \beta}{w_n} + \gamma w_n.
\end{equation}
Representing this as a system of equations in $y_n$ and $z_n$, defined as in (\ref{2:yz}). We again using the method of multiple scales by considering the slow variation in terms of $s =  \eps n$, and apply the transformation given in (\ref{1:scaling}). Hence the rescaled equations become (after Taylor expansion in $s$)
\begin{subequations}
\begin{align*}
\left[u(s,\eps) + \sum_{j=0}^{\infty} \frac{2\eps^{2j} v^{(2j)}(s,\eps)}{(2j)!} \right]u(s,\eps) &= \alpha s + \eps \beta + \gamma u(s,\eps)^2,\\
\left[v(s,\eps) + \sum_{j=0}^{\infty} \frac{2\eps^{2j} u^{(2j)}(s,\eps)}{(2j)!} \right]v(s,\eps) &= \alpha s + \eps \beta + \gamma v(s,\eps)^2. 
\end{align*}
\end{subequations}
We expand the solutions $u(s,\eps)$ and $v(s,\eps)$ as a formal series in powers of $\eps$, rather than half-integer powers as in section \ref{dPI_geq0}, as there is no $\eps^{1/2}$ present in the rescaled equations. We denote the coefficients multiplying $\eps^m$ as $u_m$ and $v_m$. As in the analysis of dP$_{\mathrm{I}}$, there are two sets of solutions, associated with the leading-order behaviours
\begin{equation*}
u_0 = -v_0 = \pm \sqrt{-\frac{\alpha s}{1+\gamma}}, \qquad u_0 = v_0 = \pm \sqrt{\frac{\alpha s}{3-\gamma}},
\end{equation*}
which we again refer to as Type A and Type B solutions respectively. We find that the general relation for higher-order terms becomes
\begin{subequations}
\begin{alignat}{2}\label{A:orderna}
\mathcal{O}(\eps^{m}):& \qquad &\sum_{j=0}^{m} u_j u_{m-j} 
                       + \sum_{j=0}^{{\lfloor}m/2{\rfloor}} \frac{2}{(2j)!}\sum_{r=0}^{m-2j} u_r v^{(2j)}_{m-r-2j}& = \gamma \sum_{j=0}^{m} u_j u_{m-j},\\
											& & \sum_{j=0}^{m} v_j v_{m-j} 
                       + \sum_{j=0}^{{\lfloor}m/2{\rfloor}} \frac{2}{(2j)!}\sum_{r=0}^{m-2j} v_r u^{(2j)}_{m-r-2j}& = \gamma \sum_{j=0}^{m} v_j v_{m-j}. 
											\label{A:ordernb}
\end{alignat}
\end{subequations}

We again define a late-order ansatz in the limit $n \rightarrow \infty$, however due to the different form of the asymptotic series, we instead select the ansatz to have the form
\begin{equation}\label{A:ansatz}
u_{m} \sim \frac{U(s)\Gamma(m+k)}{\chi(s)^{m+k}},\qquad v_{m} \sim \frac{V(s)\Gamma(m+k)}{\chi(s)^{m+k}}.
\end{equation}

Applying this ansatz to (\ref{A:orderna})-(\ref{A:ordernb}) and matching leading orders of $m$ as $m \rightarrow \infty$ gives new singulant and prefactor equations. The singulant equations for Type A and Type B solutions respectively are now given by
\begin{equation*}
\textrm{Type A:}\quad\cosh(-\chi') = \gamma,\qquad \textrm{Type B:}\quad2+\cosh(-\chi') = \gamma,
\end{equation*}
with the condition that $\chi(0)=0$ in both cases. By considering higher orders in $m$ as $m \rightarrow \infty$, we may obtain new prefactor equations, and perform a Stokes switching analysis in a similar fashion to the analysis performed on dP$_{\mathrm{I}}$.

We therefore see that, although the exponential asymptotic method described in this study may be used to study the Stokes-switching behaviour of integrable equations such as the discrete Painlev\'e equations, it does not depend on the integrability of the problem, and hence may be applied to a much more broad class of equations without significant changes to the process.

\end{document}